\documentclass[conf]{new-aiaa}
\usepackage[utf8]{inputenc}

\usepackage{graphicx}
\usepackage{amsmath}
\usepackage[version=4]{mhchem}
\usepackage{siunitx}
\usepackage{longtable,tabularx}
\usepackage{listings}
\usepackage{epsfig}
\usepackage{amsmath}
\usepackage{amsfonts}
\usepackage{amsbsy}
\usepackage{mathdots}
\usepackage{mathdots}
\usepackage{lipsum}
\usepackage{enumerate}
\usepackage{amsmath}
\usepackage[mathscr]{euscript}
\usepackage{url}
\usepackage{hyphenat}
\usepackage{verbatim}
\usepackage{accents}
\usepackage{graphicx}
\usepackage{subcaption}
\usepackage{listings}

\usepackage{graphicx}
\usepackage{grffile}
\usepackage{algorithm,algorithmicx}
\usepackage{algpseudocode}
\algnewcommand\algorithmicinput{\textbf{Input: }}
\algnewcommand\algorithmicoutput{\textbf{Output: }}
\usepackage{afterpage}
\usepackage{epstopdf}
\usepackage{lineno,hyperref}
\usepackage{smartdiagram} 
\usesmartdiagramlibrary{additions}
\usetikzlibrary{shapes.geometric,arrows,quotes}
\usepackage{tikz}
\usepackage{pgfplots}

\newcommand{\hwplotA}{\raisebox{2pt}{\tikz{\draw[red,solid,line width=0.9pt](0,0) -- (5mm,0);}}}
\newcommand{\hwplotB}{\raisebox{2pt}{\tikz{\draw[black!60!green,dotted,line width=1.2pt](0,0) -- (5mm,0);}}}
\newcommand\sbullet[1][.5]{\mathbin{\vcenter{\hbox{\scalebox{#1}{$\bullet$}}}}}

\usepackage[normalem]{ulem}

\newif\ifdebug
\debugtrue 

\usepackage[utf8]{inputenc}
\usepackage{amsfonts, amsmath}
\usepackage{graphics}
    \graphicspath{{fig/}, {ext/}, {dat/}}
\usepackage{tikz}
	\usetikzlibrary{external, positioning, arrows.meta, calc, decorations.markings}
	\usetikzlibrary{decorations.markings} 
	\tikzexternaldisable
\usepackage{pgfplots}
	\pgfplotsset{compat=newest}
	\usepgfplotslibrary{colorbrewer, statistics, groupplots, fillbetween}
\usepackage{pgfplotstable}
	\newlength{\figurewidth}
	\newlength{\figureheight}
\usepackage{booktabs}
\usepackage[capitalise, noabbrev, nameinlink]{cleveref}
\usepackage{autonum} 
\usepackage{tcolorbox}
    \tcbuselibrary{most}
\usepackage{siunitx}

\ifdebug
\usepackage{lipsum}
\fi

\DeclareSymbolFont{bbold}{U}{bbold}{m}{n}
\DeclareSymbolFontAlphabet{\mathbbold}{bbold}

\tikzset{%
	>={Stealth[length=2mm, width=1.75mm]},
	default line/.style={%
		thick,
		line cap=round,
	},
	default dashed line/.style={%
		default line,
		dashed,
		mark options={solid}
	},
	default markers/.style={%
		mark=*,
		mark size=1.5pt,
	},=
}

\pgfkeys{%
    /pgf/number format/.cd,1000 sep={}
}

\pgfplotsset{%
	default axis/.style={%
		width=\figurewidth,
		height=\figureheight,
		major tick length={2pt},
		minor tick length={2pt},
		every tick/.style={black, line cap=round},
		ticklabel style={font=\scriptsize},
		legend style={%
		    draw=none,
		    font=\scriptsize,
		    at={(1.03, 1)},
		    anchor=north west,
		    fill=none,
		    legend cell align=left
		},
		cycle list/Set1,
		axis on top,
	},
	default error bar/.style={%
		only marks,
		mark size=1.25pt,
		line cap=round,
		error bars/.cd,
		y dir=both,
		y explicit,
	}
}

\tcbset{%
	outer arc=2pt,
	arc=1pt,
	attach boxed title to top left={yshift=-3mm, xshift=3mm},
	enhanced,
	boxed title style={
		top=3pt,
		bottom=3pt,
	},
  	fonttitle=\sffamily,
	top=12pt,
}

\newtcolorbox[auto counter, number within=section]{todobox}[1][]{%
	colframe=NavyBlue,
	colback=NavyBlue!20,
	boxed title style={
		colback=NavyBlue,
	},
  	title=TO DO,
}

\newtcolorbox[auto counter, number within=section]{infobox}[1][]{%
	colframe=OliveGreen,
	colback=OliveGreen!20,
	boxed title style={
		colback=OliveGreen,
	},
  	title=INFO,
}

\colorlet{old}{white!90!black}

\pgfplotsset{%
	index set/.style={%
		width=\figurewidth,
		height=\figureheight,
        xmin={-0.1},
        ymin={-0.1},
		major tick length={0pt},
        axis line style={{ultra thin, draw opacity=0}},
        axis equal,
        xlabel={$\ell_1$},
        ylabel={$\ell_2$},
        clip=false
    },
    3 by 3/.style={%
        xmax={4.1},
        ymax={4.1},
        xlabel={$\ell_1$},
        xtick={0.5, 1.5, 2.5, 3.5},
        xticklabels={0, 1, 2, 3},
        ytick={0.5, 1.5, 2.5, 3.5},
        yticklabels={0, 1, 2, 3},
        ylabel={$\ell_2$},
    },
    4 by 3/.style={%
        xmax={5.1},
        ymax={4.1},
        xlabel={$\ell_1$},
        xtick={0.5, 1.5, 2.5, 3.5, 4.5},
        xticklabels={0, 1, 2, 3, 4},
        ytick={0.5, 1.5, 2.5, 3.5},
        yticklabels={0, 1, 2, 3},
        ylabel={$\ell_2$},
    }
}

\everymath{\displaystyle}

\title{The multifaceted nature of uncertainty in structure-property linkage with crystal plasticity finite element model}

\author{Anh Tran\footnote{Technical Staff Member, Scientific Machine Learning.}}
\affil{Sandia National Laboratories, Albuquerque NM 87123.}
\author{Pieterjan Robbe\footnote{Postdoctoral Appointee, Plasma \& Reacting Flow Science.}}
\affil{Sandia National Laboratories, Livermore, CA 94550.}
\author{Tim Wildey\footnote{Technical Staff Member, Scientific Machine Learning.}}
\affil{Sandia National Laboratories, Albuquerque NM 87123.}
\author{David Montes de Oca Zapiain\footnote{Technical Staff Member, Computational Materials \& Data Science.}}
\affil{Sandia National Laboratories, Albuquerque NM 87123.}
\author{Hojun Lim\footnote{Technical Staff Member, Computational Materials \& Data Science.}}
\affil{Sandia National Laboratories, Albuquerque NM 87123.}

\begin{document}

\maketitle

\begin{abstract}

Uncertainty quantification (UQ) plays a critical role in verifying and validating forward integrated computational materials engineering (ICME) models. Among numerous ICME models, the crystal plasticity finite element method (CPFEM) is a powerful tool that enables one to assess microstructure-sensitive behaviors and thus, bridge material structure to performance. Nevertheless, given its nature of constitutive model form and the randomness of microstructures, CPFEM is exposed to both aleatory uncertainty (microstructural variability), as well as epistemic uncertainty (parametric and model-form error). Therefore, the observations are often corrupted by the microstructure-induced uncertainty, as well as the ICME approximation and numerical errors. In this work, we highlight several ongoing research topics in UQ, optimization, and machine learning applications for CPFEM to efficiently solve forward and inverse problems.

The first aspect of this work addresses the UQ of constitutive models for epistemic uncertainty, including both phenomenological and dislocation-density-based constitutive models, where the quantities of interest (QoIs) are related to the initial yield behaviors.
We apply a stochastic collocation (SC) method to quantify the uncertainty of the three most commonly used constitutive models in CPFEM, namely phenomenological models (with and without twinning), and dislocation-density-based constitutive models, for three different types of crystal structures, namely face-centered cubic (fcc) copper (Cu), body-centered cubic (bcc) tungsten (W), and hexagonal close packing (hcp) magnesium (Mg).

The second aspect of this work addresses the aleatory and epistemic uncertainty with multiple mesh resolutions and multiple constitutive models by the multi-index Monte Carlo method, where the QoI is also related to homogenized materials properties.
We present a unified approach that accounts for various fidelity parameters, such as mesh resolutions, integration time-steps, and constitutive models simultaneously. We illustrate how multilevel sampling methods, such as multilevel Monte Carlo (MLMC) and multi-index Monte Carlo (MIMC), can be applied to assess the impact of variations in the microstructure of polycrystalline materials on the predictions of macroscopic mechanical properties.

The third aspect of this work addresses the crystallographic texture study of a single void in a cube. Using a parametric reduced-order model (also known as parametric proper orthogonal decomposition) with a global orthonormal basis as a model reduction technique, we demonstrate that the localized dynamic stress and strain fields can be predicted as a spatiotemporal problem.

The fourth aspect of this work highlights the constitutive model calibration using an optimization under microstructure-induced uncertainty with Bayesian optimization. To account for natural variability or the aleatory uncertainty of microstructure, we average the loss function over an ensemble of microstructures and couple the Monte Carlo estimator with an asynchronous parallel Bayesian optimization to calibrate a phenomenological constitutive model. The framework is demonstrated for 304L stainless steel.

The fifth aspect of this work solves a stochastic inverse problem in the structure-property relationship. In this aspect, we seek to consistently learn a distribution of microstructure features, in the sense that the forward propagation of this microstructure feature distribution through CPFEM matches a target distribution of homogenized materials properties. 

\end{abstract}



\section{Introduction}

Process-structure-property relationship is the hallmark of materials science across multiple length-scales and time-scales. Along with experimental materials science, numerous integrated computational materials engineering (ICME) models have been developed over the last several decades to accurately predict and reliably quantify uncertainty for the prediction. 
The computational ICME approach and the emerging physics-informed and physics-constrained machine learning (ML) in materials science have significantly accelerated the materials design process to tailor materials properties depending on the need~\cite{de2019new}. 
For materials science, uncertainty quantification (UQ) is an essential part, since microstructures are inherently noisy and can only be captured by statistical microstructure descriptors. 
Optimization (Opt) also plays an important role, mainly in calibrating ICME models to establish a predictive science process. 
The computer predictions with quantified uncertainty have long been visioned~\cite{oden2010computer1,oden2010computer2}, and it still holds true for ICME in computational materials science. 

For structure-property relationship, phase-field and crystal plasticity finite element model (CPFEM) are arguably the most successful computational tools to numerically investigate different materials phenomena. 
In this paper, we highlight several on-going research efforts in UQ, optimization, and ML applications for CPFEM. The numerical implementations are mainly done through DREAM.3D~\cite{groeber2014dream} and DAMASK~\cite{roters2012damask,roters2019damask}. The first code coupling was demonstrated in~\cite{diehl2017identifying}.

The remaining of the paper is organized as follows. 
Section~\ref{sec:uqConstvModel} quantifies uncertainty with respect to three different constitutive models for initial yield behaviors. 
Section~\ref{sec:mimc-cpfem} applied multi-level/multi-index Monte Carlo (MLMC/MIMC) approach to quantify uncertainty for homogenized materials properties. 
Section~\ref{sec:ROM} develops a parametric reduced-order model using proper-orthogonal decomposition method to emulate the full-field stress-strain response of a void model. 
Section~\ref{sec:ConstvMatCal} demonstrates the asynchronous parallel Bayesian optimization to calibrate a phenomenological constitutive model for stainless steel 304L. 
Section~\ref{sec:StochInv} applies a data-consistent stochastic inverse UQ technique to infer a distribution of microstructure features such that the forward propagation through CPFEM matches a target distribution of homogenized materials properties. 
Section~\ref{sec:DiscConcl} discusses and concludes the paper.

\section{UQ of constitutive models in CPFEM}
\label{sec:uqConstvModel}

In this section, we highlight our recent effort~\cite{tran2022microstructure} in quantifying \textit{epistemic} uncertainty associated with initial yield behaviors, which are characterized by yield strength $\varepsilon_\text{Y}$ and yield stress $\sigma_\text{Y}$ at 0.2\% offset. In this approach, we impose a uniform prior for all constitutive model parameters. 
Three constitutive models are considered in this study: phenomenological constitutive model \textit{with} and \textit{without} twinning, dislocation-density-based constitutive model. Interested readers are referred to the DAMASK paper~\cite{roters2019damask} (cf. Section 6) for more modeling description regarding for constitutive models. 

The generalized Wiener-Askey polynomial chaos expansion~\cite{xiu2002wiener} represents the second-order random process $f(\theta)$ as
\begin{equation}
\label{eq:gPC}
\begin{array}{lll}
f(\theta) & = & c_0 I_0 + \sum_{i_1=1}^{\infty} c_{i_1} I_1(\xi_{i_1}(\theta)) \\
&+& \sum_{i_1=1}^{\infty} \sum_{i_2=1}^{\infty} c_{i_1 i_2} I_2(\xi_{i_1}(\theta), \xi_{i_2}(\theta)) \\
&+& \sum_{i_1=1}^{\infty} \sum_{i_2=1}^{\infty} \sum_{i_3=1}^{\infty} c_{i_1 i_2 i_3} I_3(\xi_{i_1}(\theta), \xi_{i_2}(\theta), \xi_{i_3}(\theta)) + \cdots,
\end{array}
\end{equation}
where $I_n(\xi_{i_1}, \cdots, \xi_{i_n})$ denotes the Wiener-Askey polynomial chaos of order $n$ in terms of the random vector $\boldsymbol{\xi} = (\xi_{i_1}, \xi_{i_2}, \dots, \xi_{i_n})$, and $c$'s are polynomial chaos expansion coefficients to be determined. Without loss of generality, Equation~\ref{eq:gPC} can be rewritten as  
\begin{equation}
\label{eq:shortGPC}
f(\theta) = \sum_{j=0}^{\infty} \widehat{f}_j \boldsymbol{\Phi}_j(\boldsymbol{\xi(\theta)}),
\end{equation}
where there is a one-to-one correspondence between the function $I_n(\xi_{i_1}, \cdots, \xi_{i_n})$ and $\boldsymbol{\Phi}_j(\boldsymbol{\xi})$.

Table~\ref{tab:WienerAskeyScheme} describes the relationship between the types of Wiener-Askey polynomial chaos and their corresponding underlying random variables. 
For uniformly distributed variables $\boldsymbol{\xi}$ used in this paper, the Wiener-Askey scheme~\cite{xiu2002wiener} requires Legendre polynomials as the polynomial basis $\{\boldsymbol{\Phi}_j\}$.

\begin{table}[!hbtp]
\centering
\caption{Relationship between the types of Wiener-Askey polynomial chaos and their underlying random variables $\theta$}
\label{tab:WienerAskeyScheme}
\begin{tabular}{lllll} \hline
random variable & probability density function                                                     & polynomial & support range \\ \hline
Gaussian     & $\frac{1}{\sqrt{2\pi}} e^{ -\frac{\theta^2}{2} } $                                  & Hermite      & $(-\infty, \infty)$              \\
uniform      & $\frac{1}{2}$                                                                       & Legendre     & $[-1,1]$               \\
beta         & $\frac{(1-\theta)^\alpha (1+\theta)^\beta}{2^{\alpha+\beta+1} B(\alpha+1, \beta+1)} $    & Jacobi       & $[-1,1]$              \\ 
gamma        & $\frac{\theta^\alpha e^{-\theta}}{\Gamma(\alpha+1)} $                                    & Laguerre     & $[0,\infty)$             \\ \hline
\end{tabular}
\end{table}
To mitigate the curse of dimensionality, we employ stochastic collocation method~\cite{babuvska2007stochastic,nobile2008sparse,xiu2009fast} to evaluate numerical integration on Gaussian abscissas and compute the polynomial chaos expansion coefficients using Smolyak sparse grid~\cite{novak1996high,novak1997curse,novak1999simple,barthelmann2000high}. 

Following~\cite{sudret2008global}, \cite{crestaux2009polynomial}, and~\cite{saltelli2010variance}, we summarize the variance-based \textcolor{black}{global} sensitivity analysis based on Sobol' decomposition as follows. 
In the spirit of generalized polynomial chaos expansion (i.e. Equation~\ref{eq:gPC} after finite truncation), the Sobol' decomposition of $f(\boldsymbol{\xi})$ into the summands of increasing dimensions as
\begin{equation}
\begin{array}{lll}
f(\xi_1, \dots, \xi_n) &=& f_0 +  \sum_{i=1}^n  \sum_{\alpha \in \mathscr{I}_1} f_\alpha {\boldsymbol{\Phi}}(\xi_i) \\
&& + \sum_{1\leq i_1 < i_2 \leq n} \sum_{\alpha \in \mathscr{I}_{i_1 i_2}} f_\alpha \boldsymbol{\Phi}(\xi_{i_1}, \xi_{i_2}) + \cdots \\
&& + \sum_{1\leq i_1 < \cdots < i_s \leq n} \sum_{\alpha \in \mathscr{I}_{i_1, \dots, i_s}} f_\alpha \boldsymbol{\Phi}(\xi_{i_1}, \dots, \xi_{i_s})  \\
&& + \cdots + \sum_{\alpha \in \mathscr{I}_{1,2,\dots,n}} f_\alpha \boldsymbol{\Phi}(\xi_1,\dots,\xi_n).
\end{array}
\end{equation}

Given a model of the form $y = f(\xi_1, \xi_2, \dots, \xi_n)$, with $y$ as a scalar, a variance-based first order effect for a generic factor $\xi_i$ can be written as $\mathbb{V}_{\xi_i}\left[ \mathbb{E}_{\boldsymbol{\xi}_{\sim i}} \left[ y | \xi_i \right] \right]$, where $\boldsymbol{\xi}_{\sim i}$ is the vector $\boldsymbol{\xi}$ without the $i$-th element, i.e. $\boldsymbol{\xi}_{\sim i} = (\xi_1, \dots, \xi_{i-1}, \xi_{i+1}, \dots, \xi_n)$. 
The main effect sensitivity index (first-order sensitivity coefficient) is written as
\begin{equation}
\label{eq:mainSensitivityIndex}
S_i = \frac{ \mathbb{V}_{\xi_i}\left[ \mathbb{E}_{\boldsymbol{\xi}_{\sim i}} \left[ y | \xi_i \right] \right] }{ \mathbb{V}[y] }.
\end{equation}
It is relatively well-known that 
\begin{equation}
\mathbb{E} \left[ \mathbb{V} \left[ y | \boldsymbol{\xi}_{\sim i} \right] \right] + \mathbb{V} \left[ \mathbb{E} \left[ y | \boldsymbol{\xi}_{\sim i} \right] \right] = \mathbb{V}[y],
\end{equation}
and therefore, the total effect sensitivity index can be obtained as
\begin{equation}
\label{eq:totalSensitivityIndex}
T_i = \frac{ \mathbb{E} \left[ \mathbb{V} \left[ y | \boldsymbol{\xi}_{\sim i} \right] \right] }{ \mathbb{V}[y] } = 1 - \frac{ \mathbb{V} \left[ \mathbb{E} \left[ y | \boldsymbol{\xi}_{\sim i} \right] \right] }{ \mathbb{V}[y] }.
\end{equation}
For interested readers, more details and implementation are available in Tang et al.~\cite{tang2010global} and Crestaux et al.~\cite{crestaux2009polynomial}, where most of the computations are based on Monte Carlo sampling $\boldsymbol{\xi}$. 
In the context of this section, we can understand $\boldsymbol{\xi}$ as the set of parameters for the underlying constitutive model, whether it is phenomenological or dislocation-density-based, and $y$ as the quantity of interest from the CPFEM model. 

\begin{figure}[!hbtp]

\begin{subfigure}[b]{0.30\textwidth}
\centering
\includegraphics[width=\textwidth, keepaspectratio]{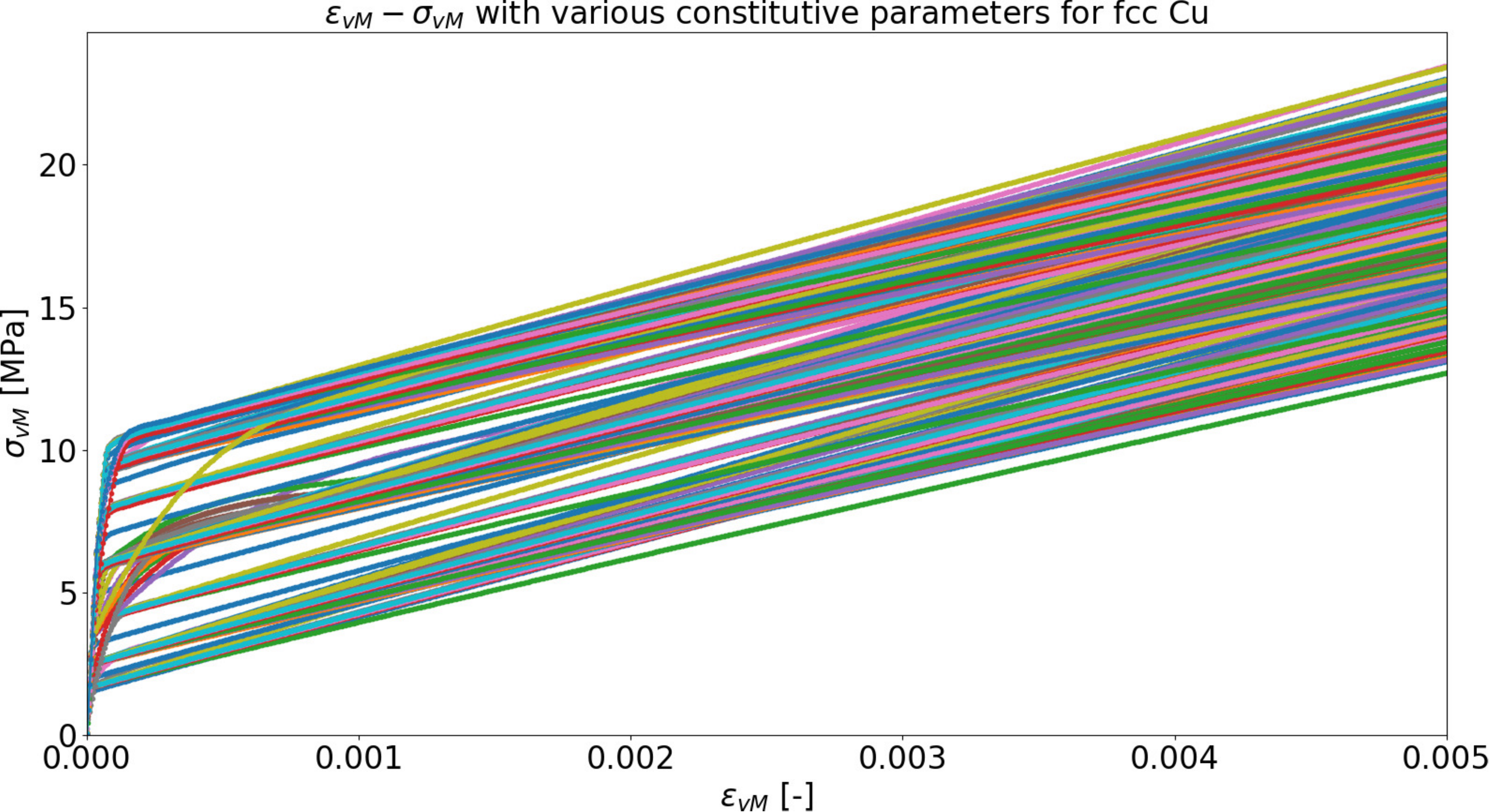}
\caption{fcc Cu.}
\label{fig:cropped_compiled-stress-strain-Cu-eps-converted-to.pdf}
\end{subfigure}
\begin{subfigure}[b]{0.30\textwidth}
\centering
\includegraphics[width=\textwidth, keepaspectratio]{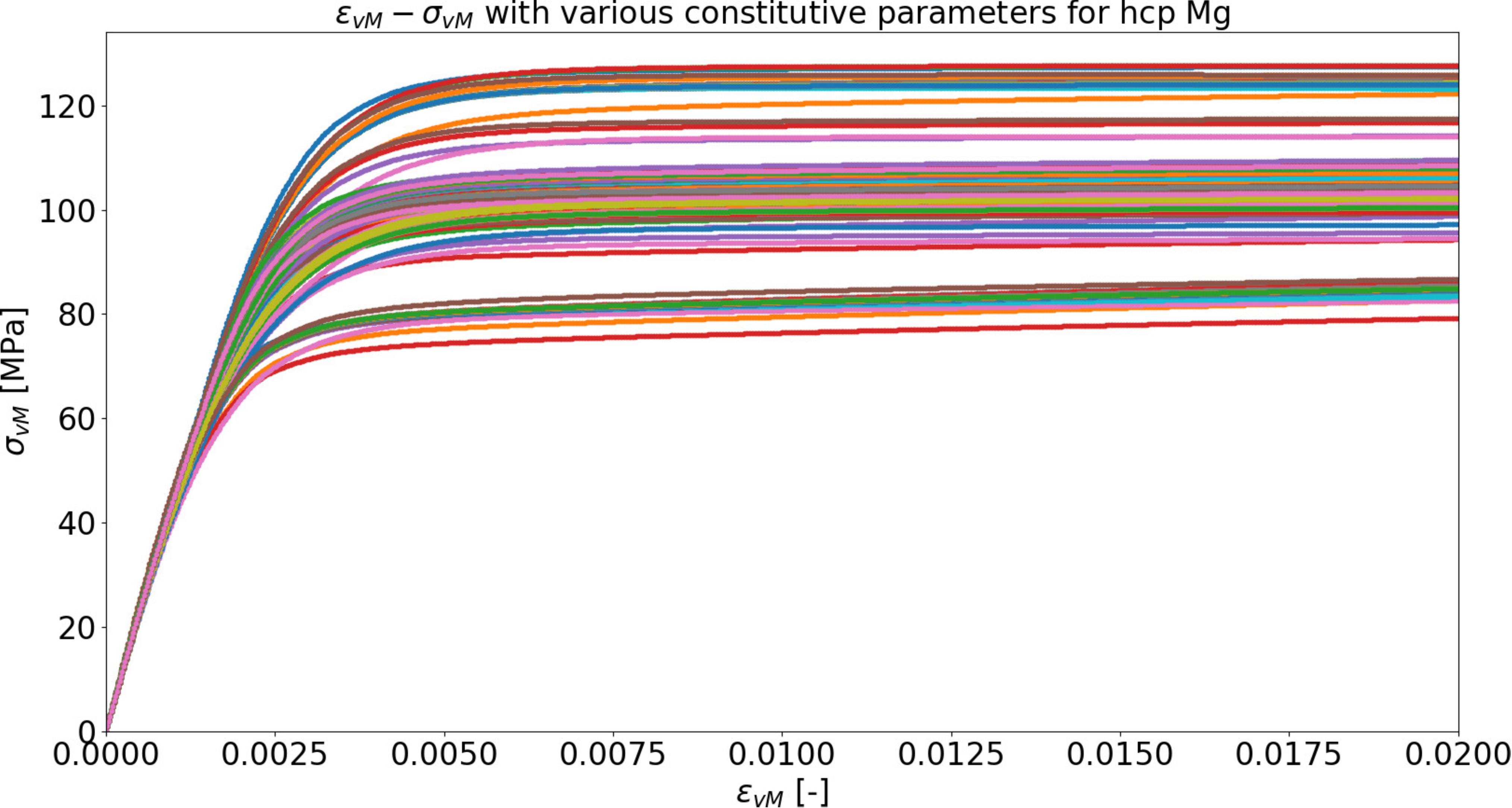}
\caption{hcp Mg.}
\label{fig:cropped_compiled-stress-strain-Mg-eps-converted-to.pdf}
\end{subfigure}
\begin{subfigure}[b]{0.30\textwidth}
\centering
\includegraphics[width=\textwidth, keepaspectratio]{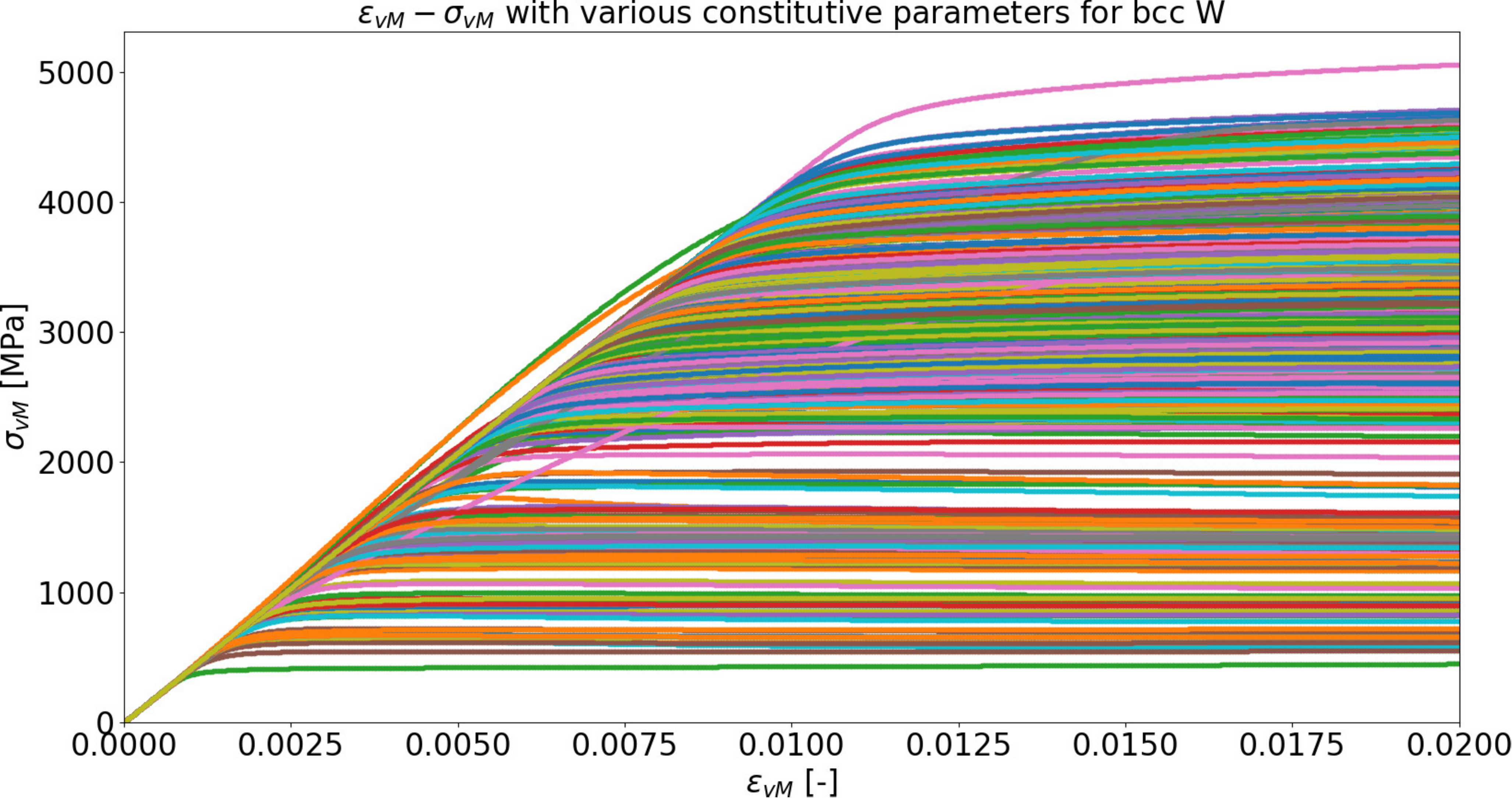}
\caption{bcc W.}
\label{fig:cropped_compiled-stress-strain-W-eps-converted-to.pdf}
\end{subfigure}
\caption{Equivalent von Mises stress-strain plots. Reprinted from~\cite{tran2022microstructure}.}
\end{figure}

\begin{figure}[!htbp]
\begin{subfigure}[b]{0.30\textwidth}
\centering
\includegraphics[width=\textwidth, keepaspectratio]{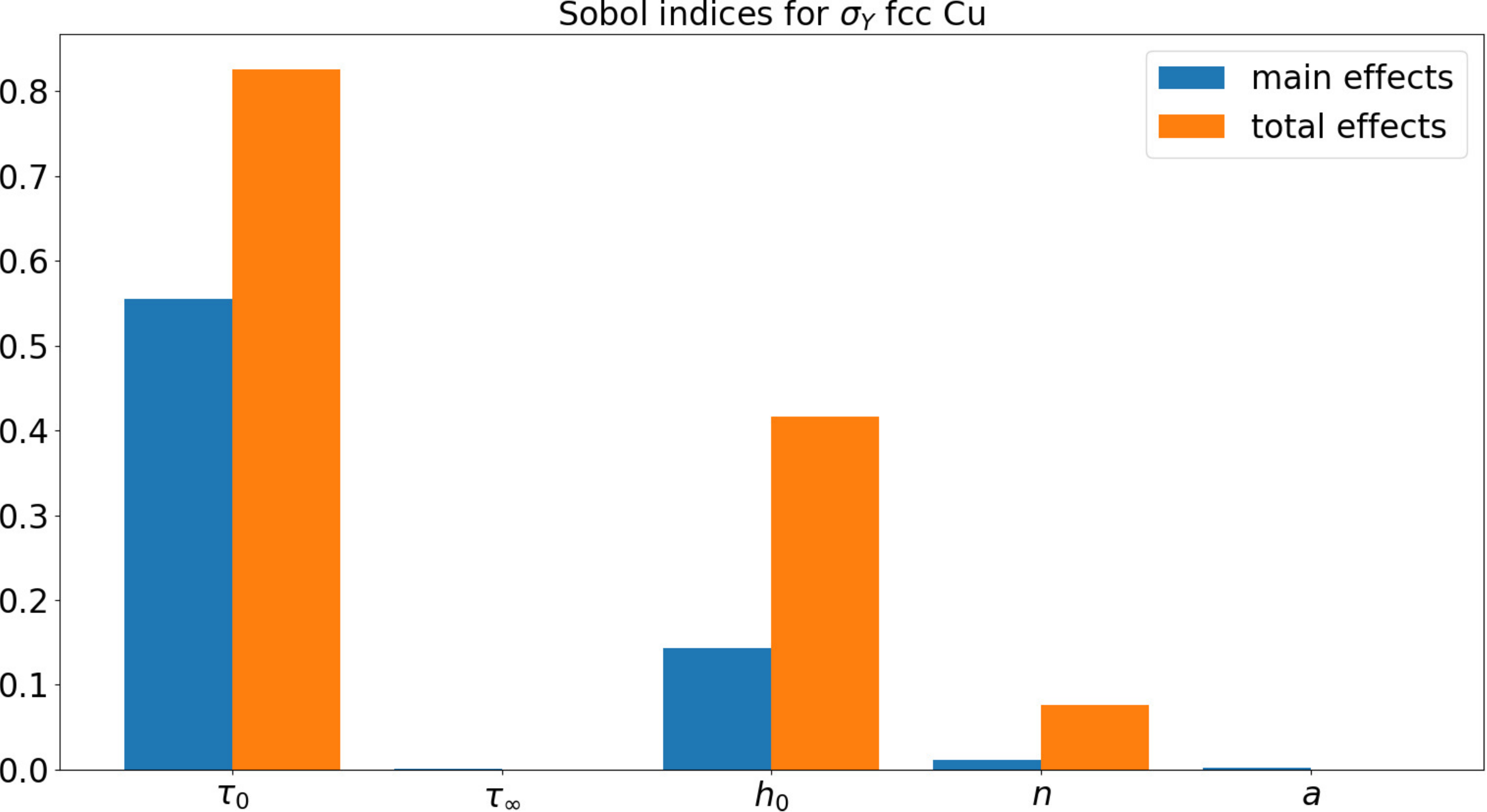}
\caption{Sobol' indices for $\sigma_{\text{Y}}$ for fcc Cu.}
\label{fig:cropped_sobol-stressYield-Cu-eps-converted-to.pdf}
\end{subfigure}
\begin{subfigure}[b]{0.30\textwidth}
\centering
\includegraphics[width=\textwidth, keepaspectratio]{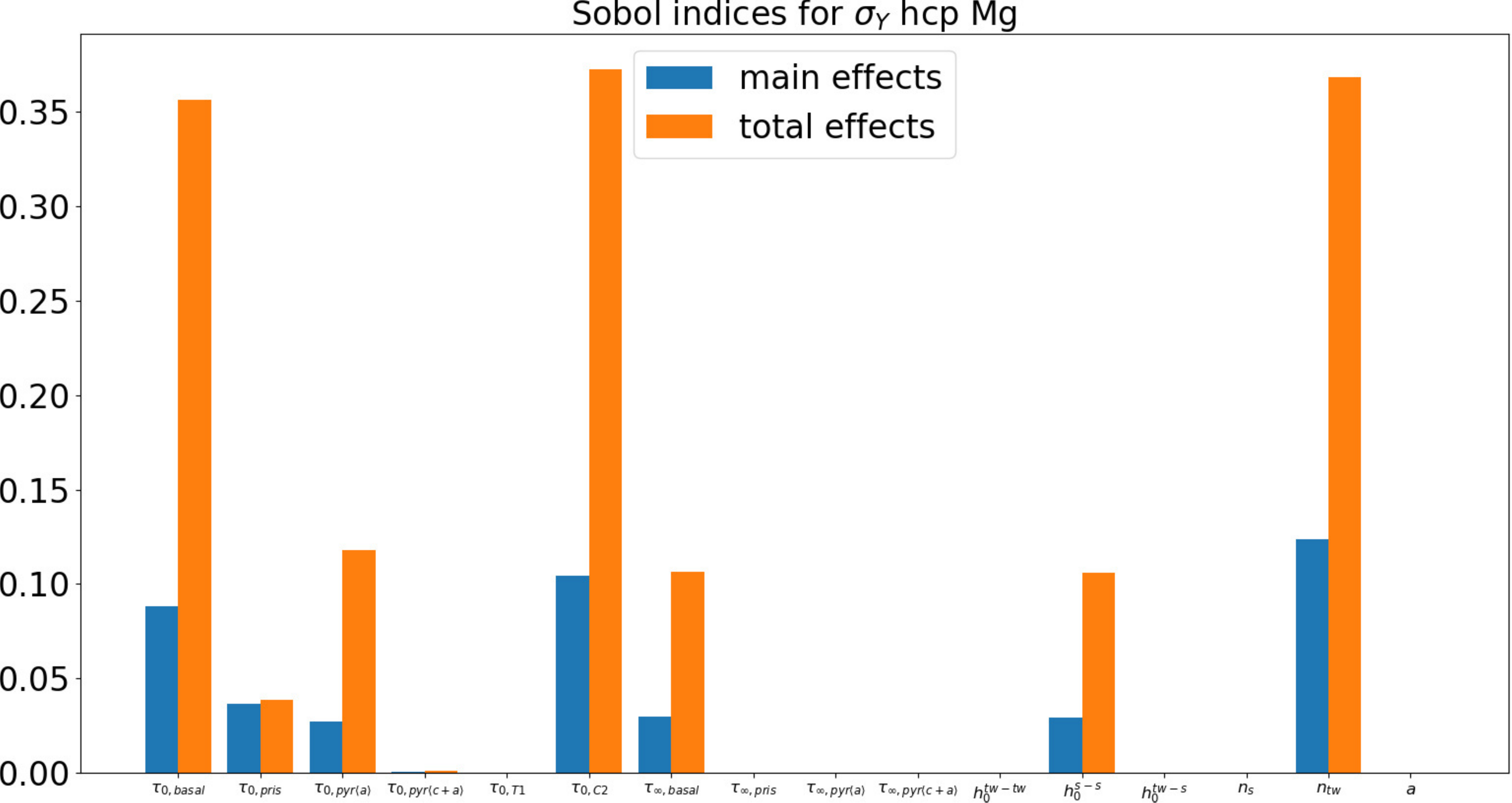}
\caption{Sobol' indices for $\varepsilon_{\text{Y}}$ for hcp Mg.}
\label{fig:cropped_sobol-stressYield-Mg-eps-converted-to.pdf}
\end{subfigure}
\begin{subfigure}[b]{0.30\textwidth}
\centering
\includegraphics[width=\textwidth, keepaspectratio]{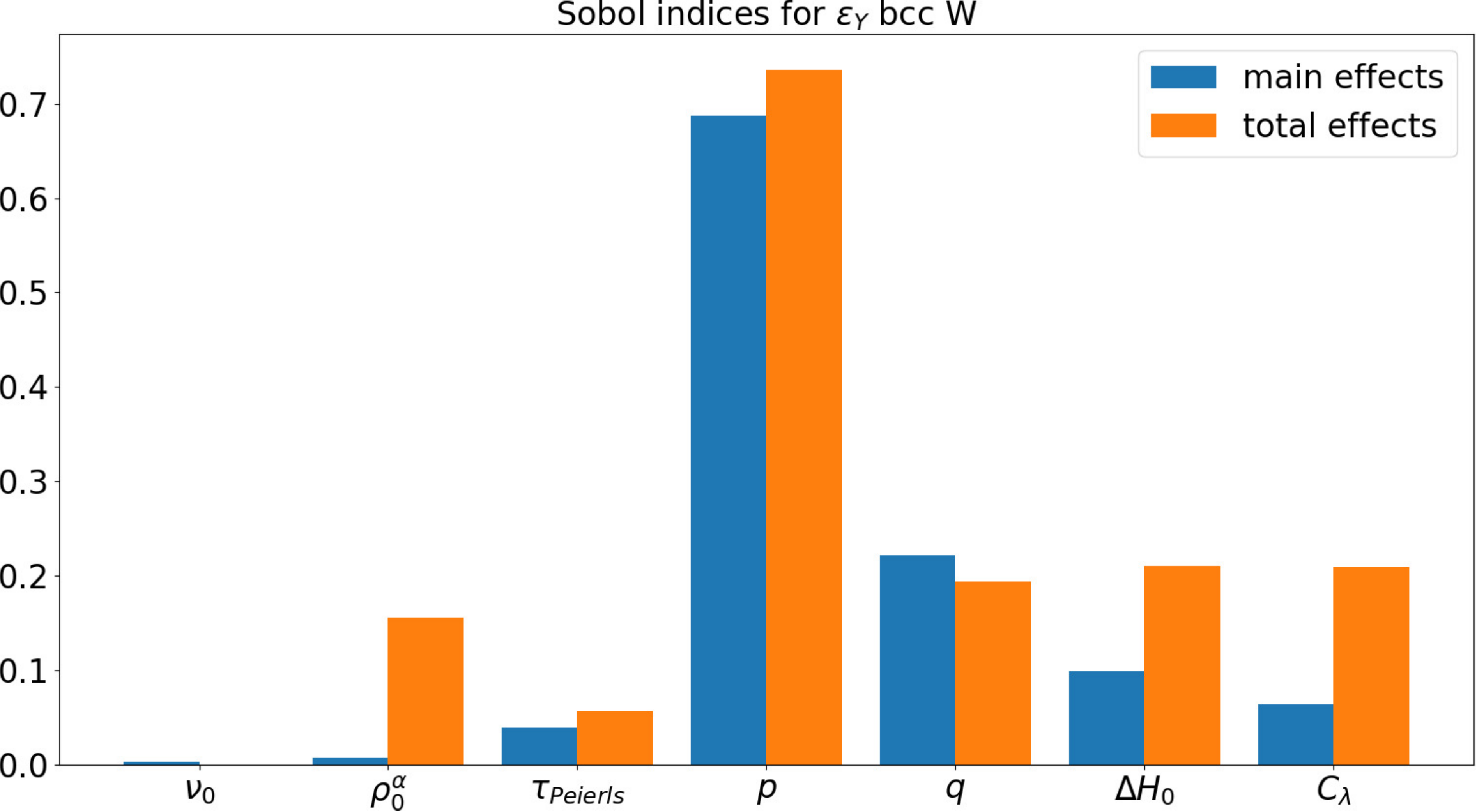}
\caption{Sobol' indices for $\varepsilon_{\text{Y}}$ for bcc W.}
\label{fig:cropped_sobol-strainYield-W-eps-converted-to.pdf}
\end{subfigure}
\caption{Global sensitivity analysis with Sobol' indices for different constitutive models. Reprinted from~\cite{tran2022microstructure}.}
\end{figure}

\section{Multi-fidelity microstructure-induced UQ for CPFEM}
\label{sec:mimc-cpfem}

The second aspect highlights our recent research effort in exploiting the well-posed fidelity hierarchy in CPFEM that is often overlooked in the literature to quantify \textit{aleatory and epistemic} uncertainty. 
The \textit{aleatory} uncertainty originates from the SERVE instantiation, whereas the \textit{epistemic} uncertainty originates from the mesh discretization for SERVEs and the plasticity constitutive model. 
In this research thrust, we apply a relatively well-known Monte Carlo, called multi-level Monte Carlo~\cite{giles2008multilevel,giles2015multilevel} and its multi-dimensional extension called multi-index Monte Carlo~\cite{haji2016multi,haji2016multi2,haji2016multi3}, to estimate a homogenized materials property in an efficient manner.

\begin{figure}[!htbp]
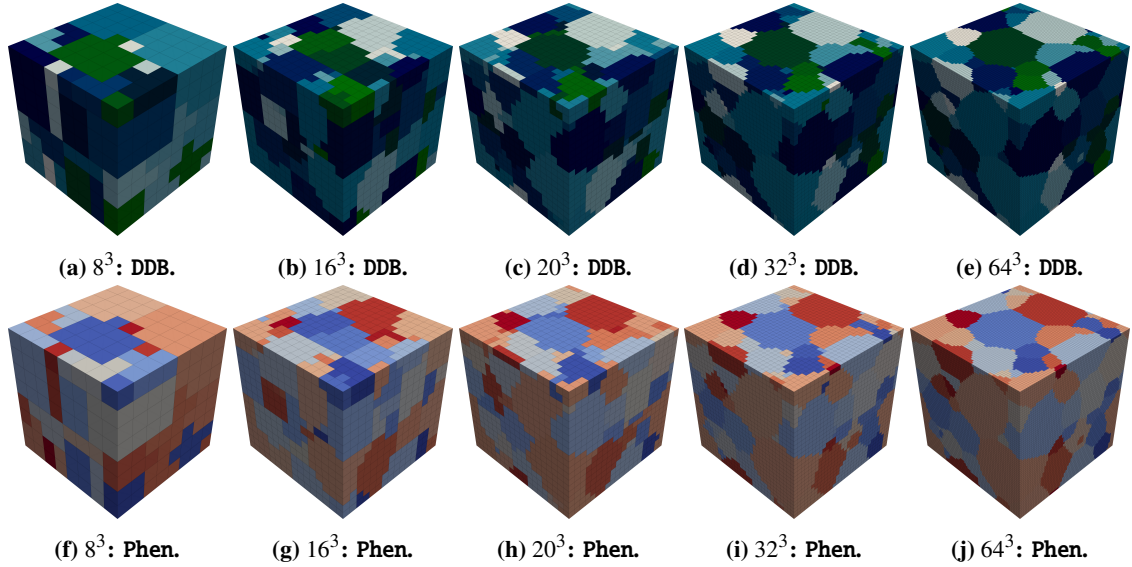

\centering
\begin{subfigure}[b]{0.175\textwidth}
\includegraphics[width=\textwidth]{cropped_single_phase_equiaxed_alphaTi_8x8x8_oceancmap-eps-converted-to.pdf}
\caption{$8^3$: \texttt{DDB}.}
\label{fig:sve8}
\end{subfigure}
\begin{subfigure}[b]{0.175\textwidth}
\includegraphics[width=\textwidth]{cropped_single_phase_equiaxed_alphaTi_16x16x16_oceancmap-eps-converted-to.pdf}
\caption{$16^3$: \texttt{DDB}.}
\end{subfigure}
\begin{subfigure}[b]{0.175\textwidth}
\includegraphics[width=\textwidth]{cropped_single_phase_equiaxed_alphaTi_20x20x20_oceancmap-eps-converted-to.pdf}
\caption{$20^3$: \texttt{DDB}.}
\end{subfigure}
\begin{subfigure}[b]{0.175\textwidth}
\includegraphics[width=\textwidth]{cropped_single_phase_equiaxed_alphaTi_32x32x32_oceancmap-eps-converted-to.pdf}
\caption{$32^3$: \texttt{DDB}.}
\end{subfigure}
\begin{subfigure}[b]{0.175\textwidth}
\includegraphics[width=\textwidth]{cropped_single_phase_equiaxed_alphaTi_64x64x64_oceancmap-eps-converted-to.pdf}
\caption{$64^3$: \texttt{DDB}.}
\label{fig:sve64}
\end{subfigure}

\begin{subfigure}[b]{0.175\textwidth}
\includegraphics[width=\textwidth]{cropped_single_phase_equiaxed_alphaTi_8x8x8-eps-converted-to.pdf}
\caption{$8^3$: \texttt{Phen}.}
\label{fig:sve8}
\end{subfigure}
\begin{subfigure}[b]{0.175\textwidth}
\includegraphics[width=\textwidth]{cropped_single_phase_equiaxed_alphaTi_16x16x16-eps-converted-to.pdf}
\caption{$16^3$: \texttt{Phen}.}
\end{subfigure}
\begin{subfigure}[b]{0.175\textwidth}
\includegraphics[width=\textwidth]{cropped_single_phase_equiaxed_alphaTi_20x20x20-eps-converted-to.pdf}
\caption{$20^3$: \texttt{Phen}.}
\end{subfigure}
\begin{subfigure}[b]{0.175\textwidth}
\includegraphics[width=\textwidth]{cropped_single_phase_equiaxed_alphaTi_32x32x32-eps-converted-to.pdf}
\caption{$32^3$: \texttt{Phen}.}
\end{subfigure}
\begin{subfigure}[b]{0.175\textwidth}
\includegraphics[width=\textwidth]{cropped_single_phase_equiaxed_alphaTi_64x64x64-eps-converted-to.pdf}
\caption{$64^3$: \texttt{Phen}.}
\label{fig:sve64}
\end{subfigure}
\caption{A schematic illustration of multi-index Monte Carlo in CPFEM with 2-dimensional fidelity level. The first fidelity variable corresponds to the mesh resolution, whereas the second fidelity variable corresponds to the constitutive plasticity model. The first fidelity variable (i.e. mesh resolution) increases from the left to the right, where as the second fidelity variable (i.e. constitutive model) increases from the bottom to the top. The low-fidelity constitutive model is phenomenological constitutive model (\texttt{Phen}) and the high-fidelity constitutive model is dislocation-density-based (\texttt{DDB}) constitutive model.}
\label{fig:MIMCSchematic}
\end{figure}

In the MLMC case study, we consider an exemplar application of statistically equivalent representative volume elements (SERVEs), where the mesh discretization could be very coarse or very fine for $\alpha$-Titanium. 
In the MIMC case study, we consider the extended version of the MLMC case study, where the first fidelity variable corresponds to the mesh resolution and the second variable corresponds to the constitutive models for an aluminum alloy. The phenomenological constitutive model is considered as the low-fidelity model, whereas the dislocation-density-based constitutive model is considered as the high-fidelity model. 
Figure~\ref{fig:MIMCSchematic} shows a schematic illustration of MIMC in two directions: the first direction -- from left to right -- corresponds to the mesh resolution, the second direction -- from bottom to top -- corresponds to the constitutive model.

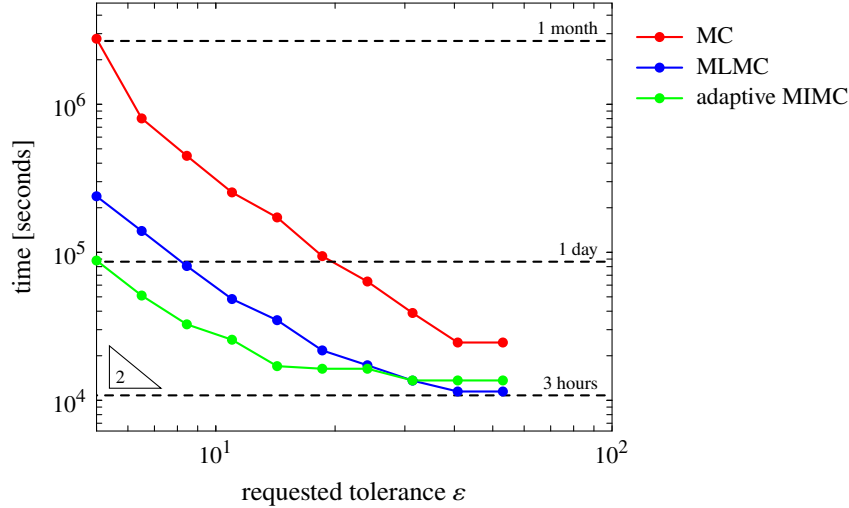
\begin{figure}[!htbp]
    \centering
    \newcommand{\LogLogSlopeTriangle}[5]
{
    \pgfplotsextra
    {
        \pgfkeysgetvalue{/pgfplots/xmin}{\xmin}
        \pgfkeysgetvalue{/pgfplots/xmax}{\xmax}
        \pgfkeysgetvalue{/pgfplots/ymin}{\ymin}
        \pgfkeysgetvalue{/pgfplots/ymax}{\ymax}

        \pgfmathsetmacro{\xArel}{#1}
        \pgfmathsetmacro{\yArel}{#3}
        \pgfmathsetmacro{\xBrel}{#1-#2}
        \pgfmathsetmacro{\yBrel}{\yArel}
        \pgfmathsetmacro{\xCrel}{\xBrel}

        \pgfmathsetmacro{\lnxB}{\xmin*(1-(#1-#2))+\xmax*(#1-#2)}
        \pgfmathsetmacro{\lnxA}{\xmin*(1-#1)+\xmax*#1}
        \pgfmathsetmacro{\lnyA}{\ymin*(1-#3)+\ymax*#3}
        \pgfmathsetmacro{\lnyC}{\lnyA+1.1*#4*(\lnxA-\lnxB)}
        \pgfmathsetmacro{\yCrel}{\lnyC-\ymin)/(\ymax-\ymin)}
        
        \coordinate (A) at (rel axis cs:\xArel,\yArel);
        \coordinate (B) at (rel axis cs:\xBrel,\yBrel);
        \coordinate (C) at (rel axis cs:\xCrel,\yCrel);

        \draw[#5]   (A)--node[pos=0.9,yshift=1ex,xshift=0.5ex] {\scriptsize #4}
                    (B)--
                    (C)-- 
                    cycle;
    }
}

\begin{tikzpicture}
\begin{axis}[ticklabel style={{font=\small}}, major tick length={2pt}, minor tick length={2pt}, every tick/.style={{black, line cap=round}}, axis on top, legend style={{draw=none, font=\small, at={(1.03,0.97)}, anchor=north west, fill=none, legend cell align=left, /tikz/every odd column/.append style={column sep=3pt}}}, xmode={log}, ymode={log}, xlabel={requested tolerance $\varepsilon$}, ylabel={time [seconds]}, xmin=5, xmax=100]
    \addplot[default line, default markers, color={red}]
        table[row sep={\\}]
        {
            \\
            53.022496865000015  24576  \\
            40.78653605000001  24576  \\
            31.374258500000003  38912  \\
            24.134045000000004  63488  \\
            18.564650000000004  94208  \\
            14.280500000000004  172032  \\
            10.985000000000003  253952  \\
            8.450000000000001  448512  \\
            6.5  802816  \\
            5.0  2772992  \\
        }
        ;
    \addlegendentry{MC}
    \addplot[default line, default markers, color={blue}]
        table[row sep={\\}]
        {
            \\
            53.022496865000015  11469.531924794  \\
            40.78653605000001  11469.5529716  \\
            31.374258500000003  13582.21158187  \\
            24.134045000000004  17251.33244051  \\
            18.564650000000004  21732.34314792  \\
            14.280500000000004  34787.371547038  \\
            10.985000000000003  48325.524031527  \\
            8.450000000000001  80784.768314001  \\
            6.5  139334.035372535  \\
            5.0  239055.66527285898  \\
        }
        ;
    \addlegendentry{MLMC}
    \addplot[default line, default markers, color={green}]
        table[x expr=500*\thisrowno{0}, row sep={\\}]
        {
            \\
            0.10604499373000004  13621.558665658  \\
            0.08157307210000003  13621.587615821  \\
            0.06274851700000002  13621.611827409999  \\
            0.04826809000000001  16340.256666803998  \\
            0.037129300000000004  16340.292646097998  \\
            0.028561000000000007  17017.313029325996  \\
            0.021970000000000007  25661.297862397994  \\
            0.016900000000000002  32601.750294422993  \\
            0.013000000000000001  51040.162352579995  \\
            0.01  87970.792336891  \\
        }
        ;
    \addlegendentry{adaptive MIMC}
    \LogLogSlopeTriangle{0.125}{0.1}{0.1}{2}{};
    \addplot[default dashed line]
    table[row sep={\\}]
    {
        1 86400 \\
        100 86400\\
    };
    \node[anchor=south east, inner sep=0pt] at (axis cs:92.5, 86400) {\scriptsize\strut 1 day};
    \addplot[default dashed line]
    table[row sep={\\}]
    {
        1 2.6784e6 \\
        100 2.6784e6\\
    };
    \node[anchor=south east, inner sep=0pt] at (axis cs:92.5, 2.6784e6) {\scriptsize\strut 1 month};
    \addplot[default dashed line]
    table[row sep={\\}]
    {
        1 10800 \\
        100 10800\\
    };
    \node[anchor=south east, inner sep=0pt] at (axis cs:92.5, 10800) {\scriptsize\strut 3 hours};
\end{axis}
\end{tikzpicture}
    \caption{Cost of the adaptive MIMC and MLMC methods compared to the (estimated) complexity of the single-level MC method, expressed in terms of the total simulation time in seconds, as a function of the tolerance $\varepsilon$ on the root mean-square error. For the target root mean-square error tolerance of $\varepsilon = 5$, the MLMC method is approximately $11.6$ times faster than the MC method, and the adaptive MIMC method is $2.7$ times faster than the MLMC method, resulting in a final speedup of adaptive MIMC over MC of $31.5\times$.}
    \label{fig:mlmc_cost_comparison}
\end{figure}

Figure~\ref{fig:mlmc_cost_comparison} compares the convergence behaviors of the Monte Carlo, MLMC, and adaptive MIMC estimators, respectively. For the MIMC case study, where the QoI is the effective Young modulus, with the root mean-square error $\varepsilon = 5$GPa, we demonstrate that the adaptive MIMC estimator is $2.7\times$ faster than the MLMC estimator, and the MLMC is $11.6\times$ faster than the MC estimator. By using an adaptive mesh resolution and an adaptive constitutive model, the homogenized materials properties can be estimated more efficiently and accurately with an unbiased estimator.

\section{A parametric ROM for void model}
\label{sec:ROM}

\begin{figure}
\centering
\includegraphics[width=\textwidth]{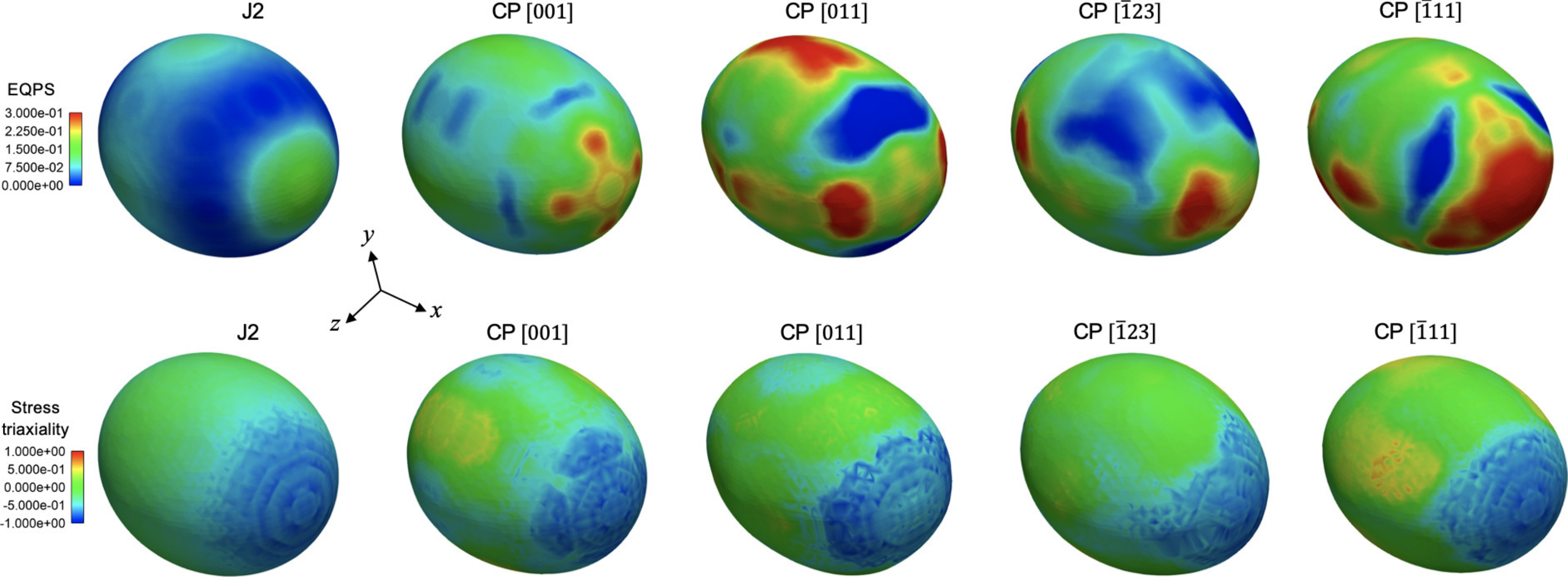}
\caption{Stress triaxiality for different crystal orientations.}
\label{fig:10triax}
\end{figure}

\begin{figure}

\begin{subfigure}[b]{0.475\textwidth}
\includegraphics[height=190px,keepaspectratio]{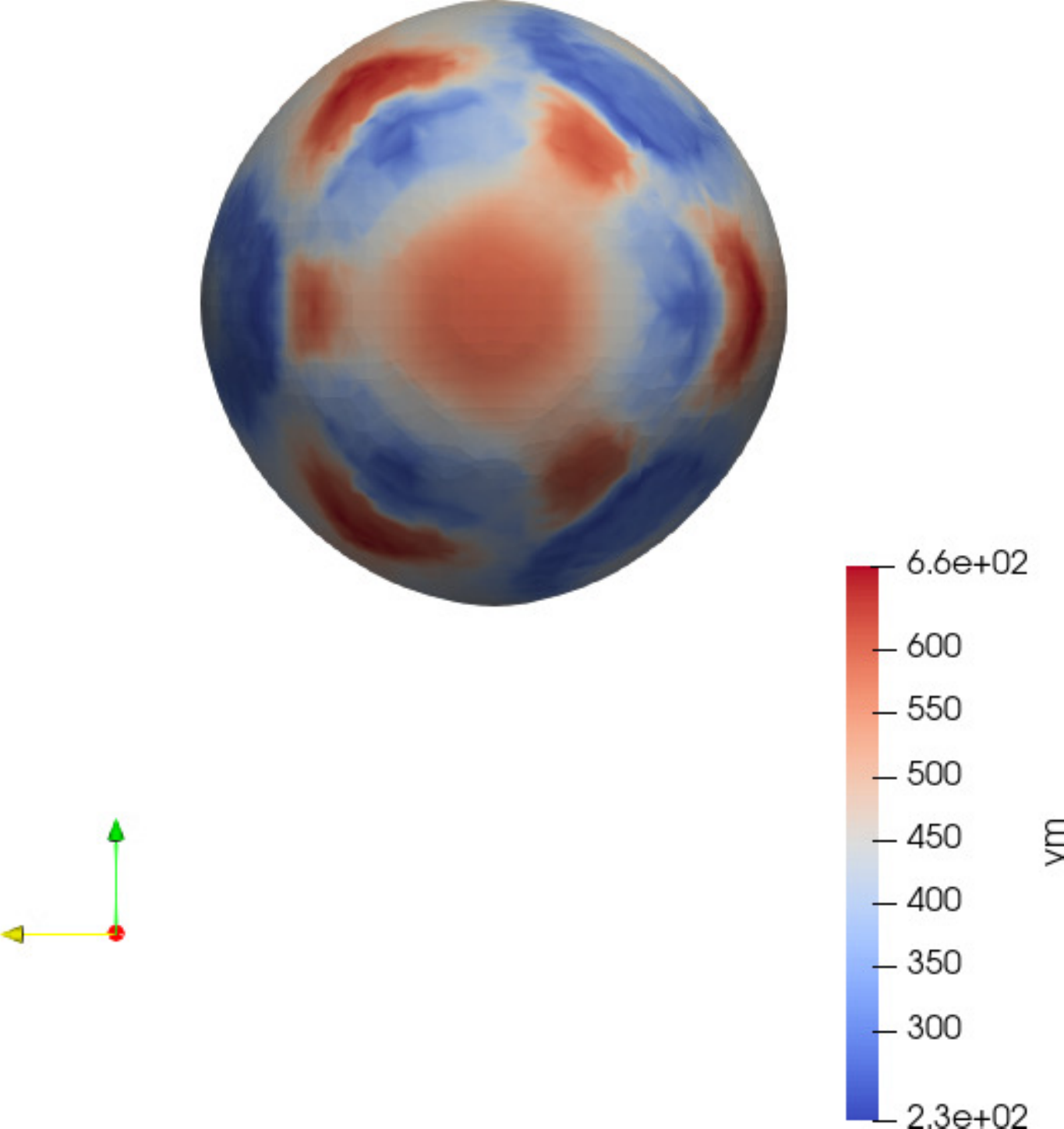}
\caption{Reduced-order model.}
\label{fig:rom}
\end{subfigure}
\hfill
\begin{subfigure}[b]{0.475\textwidth}
\centering
\includegraphics[height=190px,keepaspectratio]{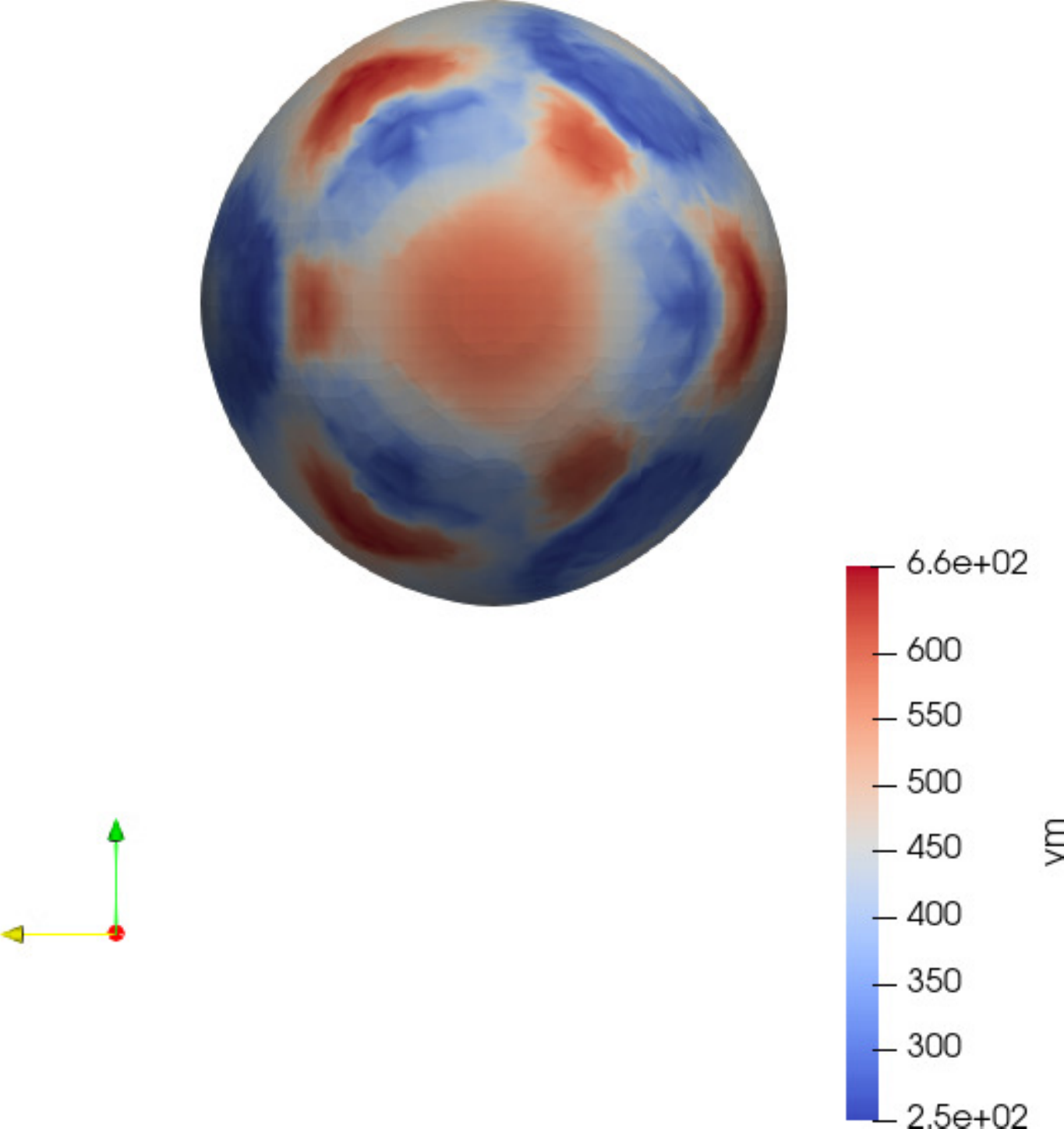}
\caption{Full-order model.}
\label{fig:fom}
\end{subfigure}
\caption{Comparison between reduced-order model (Figure~\ref{fig:rom}) and full-order model (Figure~\ref{fig:fom}) in the $[\overline{1}11]$ direction shows an excellent agreement between reduced-order model and full-order model.}
\label{fig:rom-fom}
\end{figure}

In the third aspect, we develop a projection-based reduced-order model for single crystal void model using CPFEM, mainly following~\cite{benner2015survey}. 
The idea behind is to decompose the random vector field $\mathbf{u}(\mathbf{x},t; \mathbf{p})$, in the spirit of Karhunen-Lo\`{e}ve expansion, into a set of deterministic spatial functions $\mathbf{l}_k(\mathbf{x})$ modulated by parameterized random time coefficients $\mathbf{q}(t;\mathbf{p})$ so that
\begin{equation}
\mathbf{u}(\mathbf{x},t; \mathbf{p}) = \sum_{k=1}^\infty  \mathbf{q}(t;\mathbf{p}) \mathbf{l}_k(\mathbf{x}). 
\label{eq:continousForm}
\end{equation}
Originally introduced by Sirovich~\cite{sirovich1987turbulence}, the method of snapshots consider a set of snapshots $\mathbf{u}_1, \mathbf{u}_2, \dots, \mathbf{u}_{m} \in \mathbb{R}^n$ of state solutions computed at different instants in time and different orientations $\mathbf{p}$ (parametrized either as Euler angles $(\rho_1,\Phi,\rho_2)$ or generalized spherical harmonics coefficients, which is a robust and effective way to represent textures and orientations in polycrystalline materials~\cite{bunge2013texture,kalidindi2015hierarchical}), where $\mathbf{u}_j \in \mathbb{R}^n$ denotes the $j$-th snapshot and one collects $m < n$ snapshots, where $n$ is typically large and in particular, (much) larger than the number of snapshots.
Define the snapshot matrix $\mathbf{U} \in \mathbb{R}^{n \times m}$ whose $j$th column is the snapshot $\mathbf{u}_j$. The (thin) singular decomposition of $\mathbf{U}$ is written as
\begin{equation}
\mathbf{U} = \mathbf{L} \mathbf{\Sigma} \mathbf{R}^\top,
\label{eq:thinSVD}
\end{equation}
where $\mathbf{L} \in \mathbb{R}^{n \times m}$ and $\mathbf{R} \in \mathbb{R}^{m \times m}$ are the left and right singular vectors of $\mathbf{U}$, respectively, $\mathbf{\Sigma} = \text{diag}(\sigma_1, \sigma_2, \dots, \sigma_{m}) \in \mathbb{R}^{{m} \times n}$ is a rectangular orthogonal matrix (i.e. $\mathbf{L}^\top \mathbf{L} = \mathbf{R}^\top \mathbf{R} = \mathbf{I}_m$), where $\sigma_1 \geq \sigma_2 \geq \cdots \geq \sigma_{m}$ are the singular values of $\mathbf{U}$. 
The proper orthogonal decomposition (POD) basis, $\mathbf{V}$, is chosen as the $r$ left singular vectors of $\mathbf{L}$ that correspond to the $r$ largest singular values.
The POD basis is ``optimal'' in the sense that, for an orthonormal basis of size $r$, it minimizes the least squares error of snapshot reconstruction
\begin{equation}
\min_{\mathbf{V} \in \mathbb{R}^{n \times r}} \lVert \mathbf{U} - \mathbf{V} \mathbf{V}^\top \mathbf{U} \rVert_F^2
=
\min_{\mathbf{V} \in \mathbb{R}^{n \times r}} \sum_{i=1}^{m} \lVert \mathbf{u}_i - \mathbf{V} \mathbf{V}^\top \mathbf{u}_i \rVert_F^2
=
\sum_{i=r+1}^{m} \sigma_i^2
\end{equation}
by the Eckart-Young theorem in Frobenius norm (cf. Theorem 2.4.8 and Section 2.5.2~\cite{golub2013matrix}).
It should be noted that the spatial modes of the direct POD are given by its left singular vectors $\mathbf{L}$, the temporal modes of the snapshot POD are given by its right singular vectors $\mathbf{R}$ since $\mathbf{U} \in \mathbb{R}^{n \times m}$. From Equation~\ref{eq:thinSVD}, let
\begin{equation}
\mathbf{Q} = \mathbf{\Sigma} \mathbf{R}^\top =  \mathbf{L}^\top \mathbf{U},
\end{equation}
where $\mathbf{Q} \in \mathbb{R}^{m \times m}$ then
\begin{equation}
\mathbf{U} = \mathbf{L} \mathbf{Q} = \sum_{k=1}^{m} l_k q_k,
\label{eq:discreteForm}
\end{equation}
where $l_k$ is the $k$th column of $\mathbf{L}$ matrix, $q_k$ is the $k$th row of $\mathbf{Q}$. We arrive at the discrete form of Equation~\ref{eq:continousForm} in temporal modes.
Under the reduced-order model representation, the random vector field is approximated by a truncated Karhunen-Lo\`{e}ve expansion as
\begin{equation}
\mathbf{u}(\mathbf{x},t; \mathbf{p}) \approx \tilde{\mathbf{u}}(\mathbf{x},t; \mathbf{p}) = \sum_{k=1}^r  \mathbf{q}(t;\mathbf{p}) \mathbf{l}_k(\mathbf{x}).
\end{equation}
Now that the model can be represented by
\begin{equation}
\mathbf{U} = \mathbf{L} \mathbf{Q},
\label{eq:BasisTimesCoefs}
\end{equation}
where $\mathbf{U} \in \mathbb{R}^{n \times m}, \mathbf{L} \in \mathbb{R}^{n \times m}, \mathbf{Q} \in \mathbb{R}^{m \times m}$ we can now interpolate the coefficient matrix $\mathbf{Q}$ to obtain
\begin{equation}
\mathbf{U}' = \mathbf{L} \mathbf{Q}',
\end{equation}
where $\mathbf{U}' \in \mathbb{R}^{n \times m'}, \mathbf{L}' \in \mathbb{R}^{n \times m}, \mathbf{Q}' \in \mathbb{R}^{m \times m'}$.
Following~\cite{sirovich1987turbulence,berkooz1993proper}, we construct a global basis approach to solve a parametric reduced-order model. 
One of the most common approaches in constructing the global basis matrices over the parameter space $\mathbf{p}$ is to concatenate the local basis matrices obtained by several parameter samples $\mathbf{p}_1, \dots, \mathbf{p}_K$.
Suppose that $\mathbf{Q}_1, \mathbf{Q}_2, \dots, \mathbf{Q}_K$ denote the local basis matrices corresponding to $\mathbf{p}_1, \dots, \mathbf{p}_K$, respectively, one can construct the global basis matrices $\mathbf{Q}$ using
\begin{equation}
\mathbf{Q} = [\mathbf{Q}_1, \mathbf{Q}_2, \dots, \mathbf{Q}_K],
\end{equation}
followed by an singular value decomposition (SVD) or a rank-revealing QR factorization to remove the rank-deficient components from $\mathbf{Q}$.
Here we employ the Galerkin projection method~\cite{bui2008model,carlberg2017galerkin} to obtain the coefficient from the global basis in $L_2$-norm, exploiting the orthonormal properties of the global basis.

Figure~\ref{fig:10triax} shows an exemplar of the stress triaxiality of a void using CPFEM for 9 different crystal orientations, along with a $J2$ plasticity model. 
For this anisotropic case study, the localized fields of the void (displacement in $x,y,z$ directions, $\sigma_\text{vM}$, stress triaxiality) are a function of time and Euler angles $(\rho_1,\Phi,\rho_2)$. 
Figure~\ref{fig:rom-fom} compares between a projection-based reduced-order model (ROM) and the full-order model (FOM, i.e. CPFEM) for the $[\overline{1}11]$ orientation, which shows an excellent agreement between the ROM and the FOM.

\section{A high-throughput Bayesian optimization for constitutive model calibration}
\label{sec:ConstvMatCal}

\begin{figure}
\centering
\includegraphics[width=\textwidth]{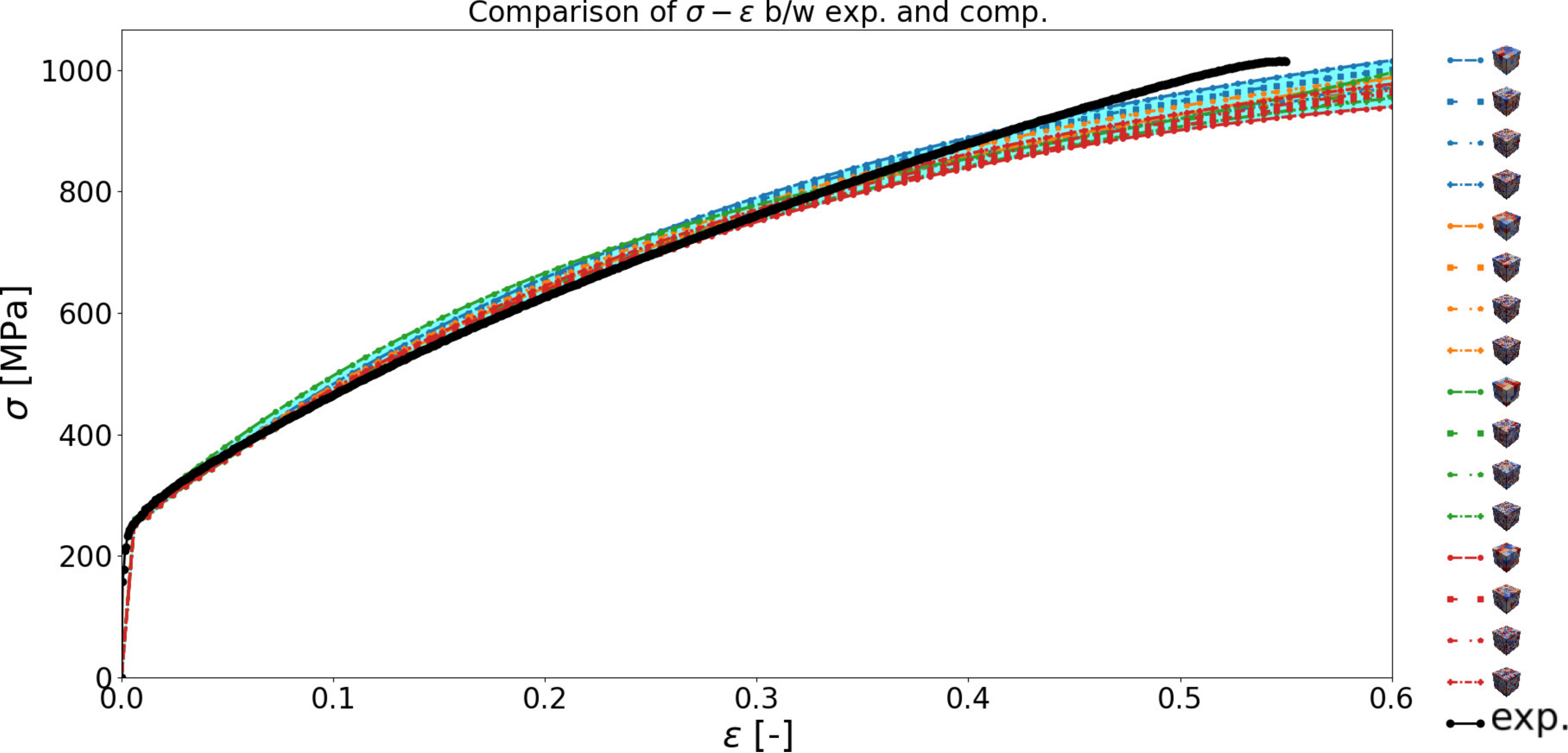}
\caption{Comparison of homogenized materials properties between experimental data and computational results for SS304L across different mesh resolutions and for different SERVEs.}
\label{fig:SS304Lopt}
\end{figure}

The fourth aspect addresses an effort in calibrating constitutive model in a high-throughput and asynchronous parallel manner, using one of our previous work~\cite{tran2022aphbo,tran2019pbo,tran2020multi}. 
In this section, we focus on the optimization under (microstructure-induced, also known as \textit{aleatory}) uncertainty with applications to constitutive model calibration. 
The loss function to be minimized is measured in $L_2$-norm and normalized by the maximum observable equivalent strain $\varepsilon_\text{vM}$. 
A set of five constitutive parameters is used to parametrize a phenomenological constitutive model for stainless steel 304L. 
At any time, 12 CPFEM simulations are performed concurrently, where the batch configuration is set as (8,4,0). 
To account for the \textit{aleatory} uncertainty, we average the loss function over an ensemble of 5 SERVEs, where the mesh of $5\times5\times5$ is used.

Figure~\ref{fig:SS304Lopt} shows the comparison between the computational results produced by the optimal 5d constitutive parameters and the experimental data (marked as solid black line). The \textit{aleatory} uncertainty is colored in a cyan shaded region (readers are referred to the color version online). 
Figure~\ref{fig:SS304Lopt} shows a good agreement between experimental data and computational results up to approximately $\varepsilon\approx 0.42$ of strain, which is considerable for CPFEM. 
The optimal constitutive parameters is found after 352 iterations using the \texttt{aphBO-2GP-3B} algorithm~\cite{tran2022aphbo}.

\section{Solving stochastic inverse in property-structure relationship with ML}
\label{sec:StochInv}

\begin{figure}[!htbp]
\begin{subfigure}[b]{0.475\textwidth}
\includegraphics[height=190px,keepaspectratio]{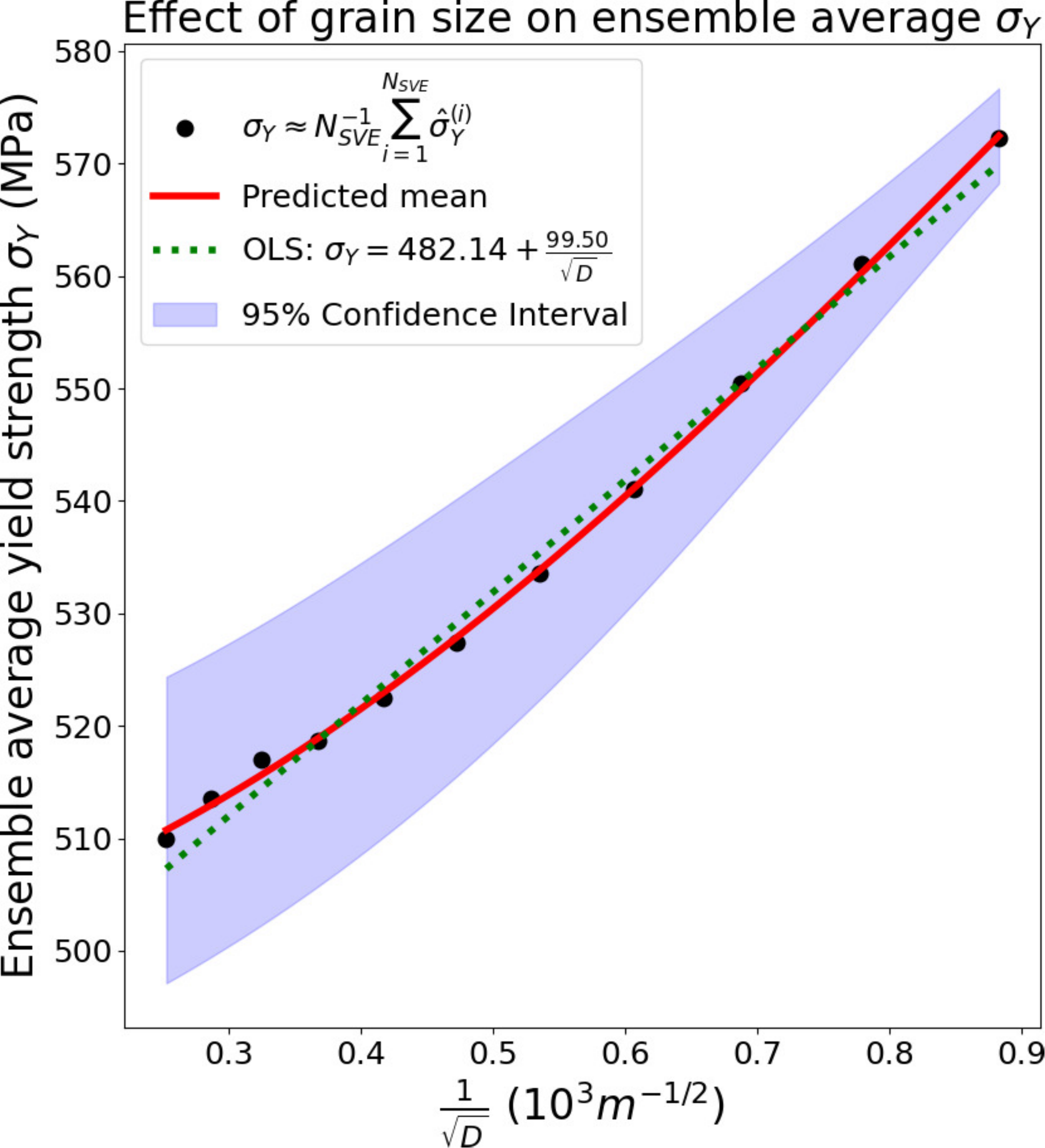}
\caption{Hall-Petch relationship by ordinary least square and Gaussian process regression.}
\label{fig:HallPetch}
\end{subfigure}
\hfill
\begin{subfigure}[b]{0.475\textwidth}
\centering
\includegraphics[height=190px,keepaspectratio]{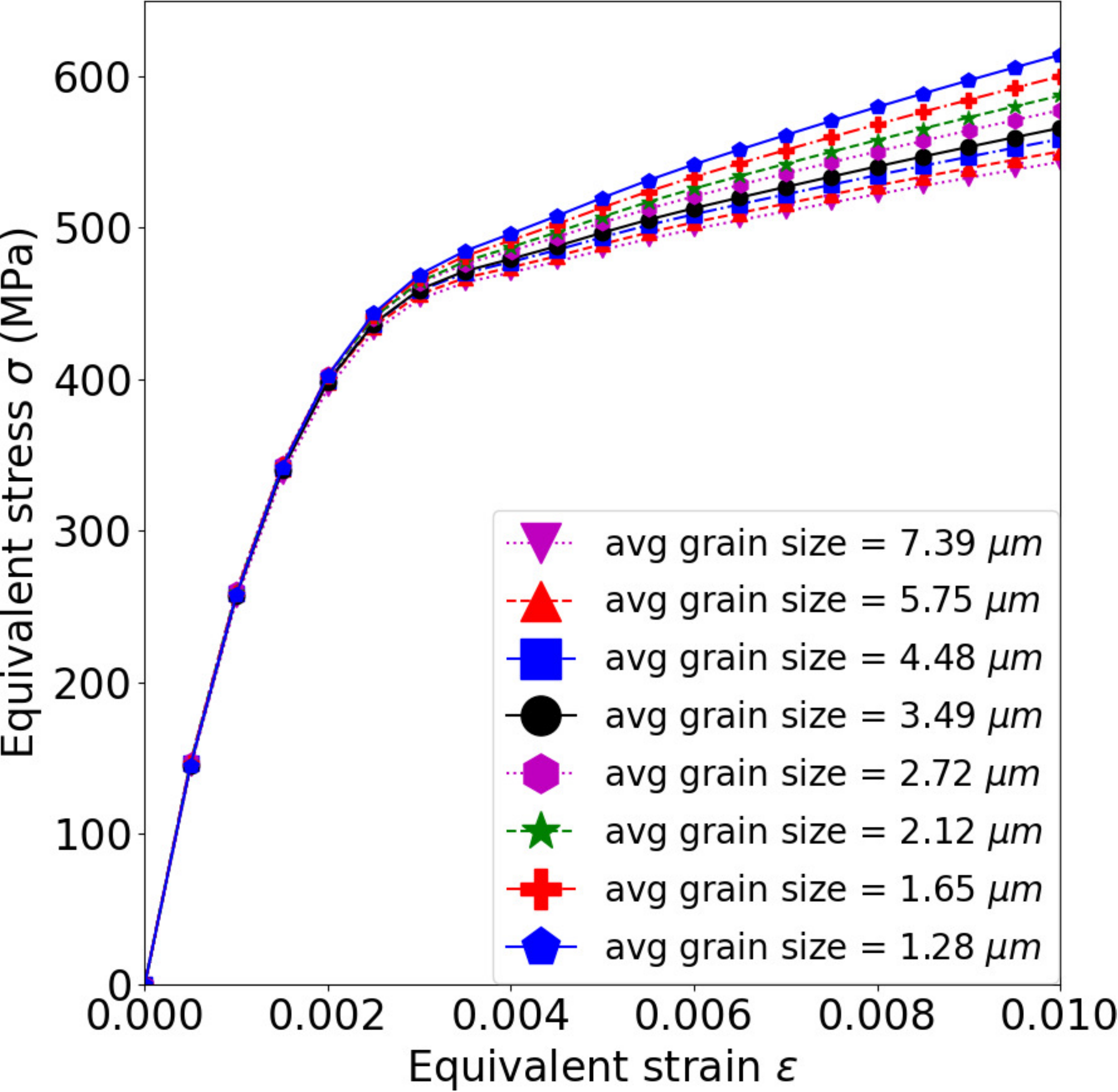}
\caption{Equivalent stress-strain curve for different average grain sizes.}
\label{fig:equivStressStrainGrainSize4}
\end{subfigure}
\caption{Figure~\ref{fig:HallPetch}: The inferred ML Hall-Petch relationship with 95\% confidence intervals for a TWIP steel, and a comparison with a least squares regression model. 
Figure~\ref{fig:equivStressStrainGrainSize4}: The representative equivalent stress-strain relationship with respect to various average grain sizes. CPFEM simulations are performed with $64\mu m \times 64\mu m \times 64\mu m$ SVE at the loading condition of $\dot{\varepsilon}_{11} = 0.001 \text{s}^{-1}$. (Reprinted with permission from~\cite{tran2020solving} \copyright 2020 The Minerals, Metals \& Materials Society.)}
\label{fig:grainsizeHallPetch_SS_TWIP}
\end{figure}

The last aspect~\cite{tran2020solving} applies a stochastic inverse UQ methodology~\cite{butler2018convergence,butler2018combining,butler2020stochastic} to infer a data-consistent distribution of microstructure features, in the sense that the push-forward distribution through CPFEM matches the target distribution of the homogenized materials properties of interests. In this case study, we employed a dislocation-density-based for twinning-induced plasticity/transformation-induced plasticity (TWIP/TRIP) Fe-22Mn-0.6C steels~\cite{wong2016crystal}, where the density of average grain size is inferred based on a ML surrogate (i.e. a heteroscedastic Gaussian process regression) for the Hall-Petch relationship to match target distribution of yield stress. 
The caveat in this case study is the notion of heteroscedastic behavior under a fixed size assumption of SERVE. 
Let $D$ denotes the average grain size. 
When the average grain size $D$ increases, the average grain volume scales as $\mathcal{O}(D^3)$, and under the assumption of fixed volume for SERVEs, the number of grain size $n$ decreases as $n \sim \mathcal{O}(D^{-3})$. The variance of the Monte Carlo estimator scales as $n^{-1} \sim \mathcal{O}(D^3)$. Hence larger average grain size would induce a larger variance in the Monte Carlo estimator. As such, the regression is heteroscedastic. 

Figure~\ref{fig:HallPetch} shows the Hall-Petch relationship obtained from CPFEM, estimated by the posterior mean of the heteroscedastic Gaussian process regression (\hwplotA) and ordinary least square (\hwplotB). The ensemble average observations are denoted as $\sbullet[.75]$. 
Figure~\ref{fig:equivStressStrainGrainSize4} shows the equivalent stress-strain curve for different average grain size at the initial yield 0.2\% offset.

\begin{figure}[!htbp]
\begin{subfigure}[b]{0.475\textwidth}
\centering
\includegraphics[height=190px,keepaspectratio]{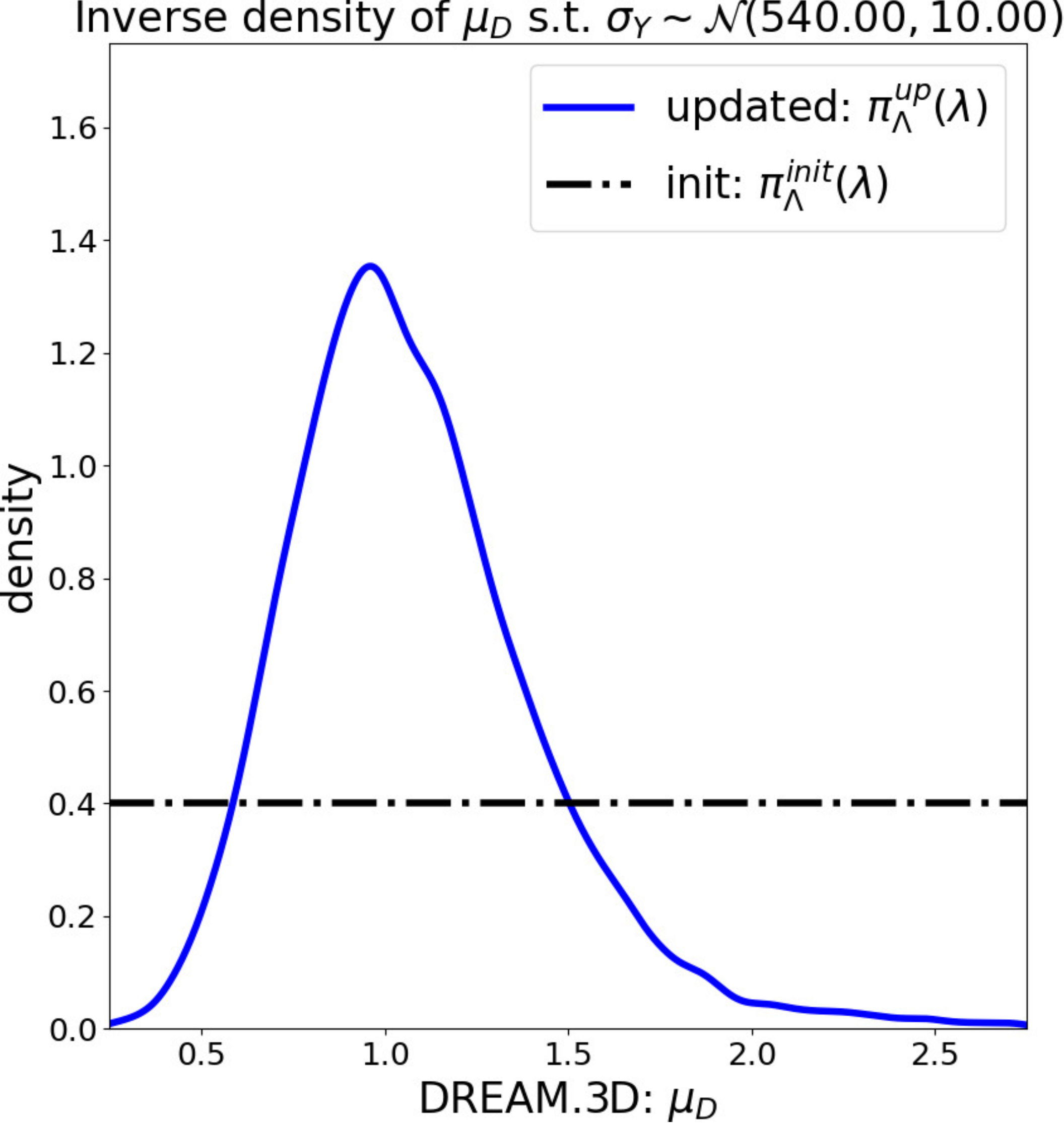}
\caption{Uniform initial density $\pi_{\mathbf{\Lambda}}^{\text{init}}(\lambda) = \mathcal{U}(0.25,2.75)$ and updated density
$\pi_{\mathbf{\Lambda}}^{\text{up}}(\lambda)$ where $\lambda = \mu_D$.}
\label{fig:posteriorNormal-540-10}
\end{subfigure}
\hfill
\begin{subfigure}[b]{0.475\textwidth}
\centering
\includegraphics[height=190px,keepaspectratio]{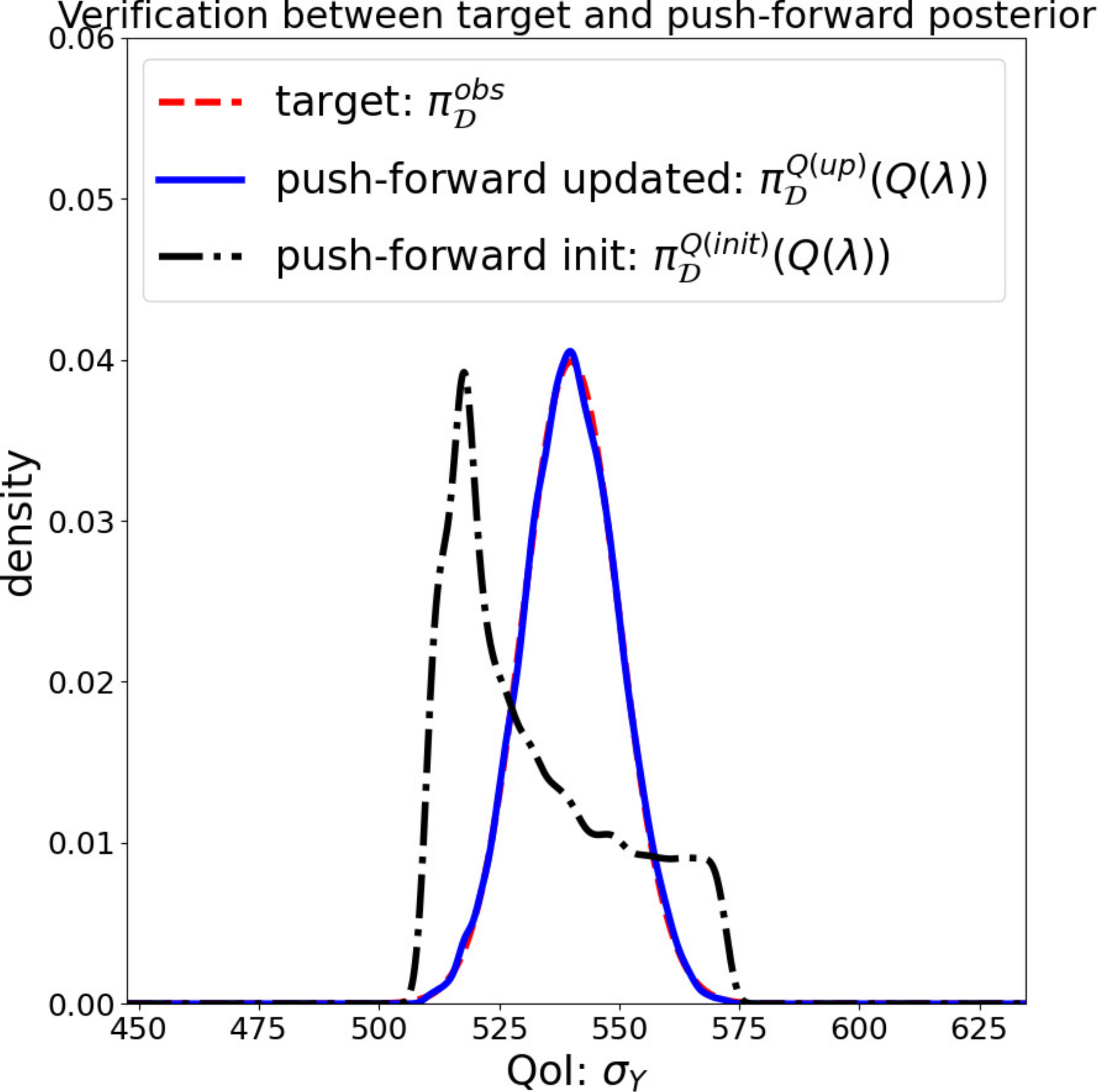}
\caption{Target $\pi_{\mathbf{\mathcal{D}}}^{\text{obs}} = \mathcal{N}(540,10)$ and push-forward of the updated density $\pi_{\mathbf{\Lambda}}^{\text{up}}(Q(\lambda))$.}
\label{fig:densityComparisonNorm-540-10}
\end{subfigure}
\caption{Figure~\ref{fig:posteriorNormal-540-10}: The initial density, $\pi_{\mathbf{\Lambda}}^{\text{init}}(\lambda) = \mathcal{U}(0.25,2.75)$, and updated density, $\pi_{\mathbf{\Lambda}}^{\text{up}}(\lambda)$, on the microstructure feature $\lambda = \mu_D$.
Figure~\ref{fig:densityComparisonNorm-540-10}: The target density on material properties, $\pi_{\mathbf{\mathcal{D}}}^{\text{obs}} = \mathcal{N} (540, 10)$, and push-forwards of the initial and updated densities. (Reprinted with permission from~\cite{tran2020solving} \copyright 2020 The Minerals, Metals \& Materials Society.)
}
\label{fig:StochInvComp}
\end{figure}

Figure~\ref{fig:posteriorNormal-540-10} shows the uniform \textit{prior} and the \textit{posterior} distributions of average grain size, where the \textit{posterior} is updated using the stochastic inverse method. 
Figure~\ref{fig:densityComparisonNorm-540-10} shows the normal target distribution $\mathcal{N}(540,10)$, the push-forward of the \textit{prior}, and the push-forward of the \textit{posterior}. 
The push-forward of the \textit{posterior} and the target distribution of the yield stress agrees very well with each other.

\section{Discussion \& Conclusion}
\label{sec:DiscConcl}

In this paper, we survey our recent or on-going research effort in UQ, optimization, and ML for structure-property relationship using CPFEM. As the microstructure is usually high-dimensional and naturally random, deploying UQ techniques is necessary to conduct an efficient and effective numerical studies. In this paper, we demonstrate multiple aspects of using UQ mathematical methodologies in the interest of CPFEM for structure-property relationship. Even though the field of UQ and multiscale computational materials science both have been found for a relatively long time, there is still plenty of research opportunities and open questions for further research. 
The objective of this paper is to highlight a few interesting and contemporary topics in UQ for CPFEM. 

\section*{Acknowledgments}
Sandia National Laboratories is a multimission laboratory managed and operated by National Technology \& Engineering Solutions of Sandia, LLC, a wholly owned subsidiary of Honeywell International Inc., for the U.S. Department of Energy’s National Nuclear Security Administration under contract DE-NA0003525.

\bibliography{lib}

\begin{thebibliography}{41}
\newcommand{\enquote}[1]{``#1''}
\providecommand{\natexlab}[1]{#1}
\providecommand{\url}[1]{\texttt{#1}}
\providecommand{\urlprefix}{URL }
\expandafter\ifx\csname urlstyle\endcsname\relax
  \providecommand{\doi}[1]{\discretionary{}{}{}https://doi.org/#1}\else
  \providecommand{\doi}[1]{\discretionary{}{}{}\urlstyle{rm}\url{https://doi.org/#1}}\fi

\bibitem[{de~Pablo et~al.(2019)de~Pablo, Jackson, Webb, Chen, Moore, Morgan,
  Jacobs, Pollock, Schlom, Toberer et~al.}]{de2019new}
de~Pablo, J.~J., Jackson, N.~E., Webb, M.~A., Chen, L.-Q., Moore, J.~E.,
  Morgan, D., Jacobs, R., Pollock, T., Schlom, D.~G., Toberer, E.~S., et~al.,
  \enquote{New frontiers for the materials genome initiative,} \emph{npj
  Computational Materials}, Vol.~5, No.~1, 2019, pp. 1--23.

\bibitem[{Oden et~al.(2010{\natexlab{a}})Oden, Moser, and
  Ghattas}]{oden2010computer1}
Oden, T., Moser, R., and Ghattas, O., \enquote{Computer predictions with
  quantified uncertainty, {P}art {I},} \emph{SIAM News}, Vol.~43, No.~9,
  2010{\natexlab{a}}, pp. 1--3.

\bibitem[{Oden et~al.(2010{\natexlab{b}})Oden, Moser, and
  Ghattas}]{oden2010computer2}
Oden, T., Moser, R., and Ghattas, O., \enquote{Computer predictions with
  quantified uncertainty, {P}art {II},} \emph{SIAM News}, Vol.~43, No.~10,
  2010{\natexlab{b}}, pp. 1--4.

\bibitem[{Groeber and Jackson(2014)}]{groeber2014dream}
Groeber, M.~A., and Jackson, M.~A., \enquote{{DREAM.3D}: a digital
  representation environment for the analysis of microstructure in {3D},}
  \emph{Integrating materials and manufacturing innovation}, Vol.~3, No.~1,
  2014, p.~5.

\bibitem[{Roters et~al.(2012)Roters, Eisenlohr, Kords, Tjahjanto, Diehl, and
  Raabe}]{roters2012damask}
Roters, F., Eisenlohr, P., Kords, C., Tjahjanto, D., Diehl, M., and Raabe, D.,
  \enquote{{DAMASK: the D{\"u}sseldorf Advanced MAterial Simulation Kit for
  studying crystal plasticity using an FE based or a spectral numerical
  solver},} \emph{Procedia Iutam}, Vol.~3, 2012, pp. 3--10.

\bibitem[{Roters et~al.(2019)Roters, Diehl, Shanthraj, Eisenlohr, Reuber, Wong,
  Maiti, Ebrahimi, Hochrainer, Fabritius et~al.}]{roters2019damask}
Roters, F., Diehl, M., Shanthraj, P., Eisenlohr, P., Reuber, C., Wong, S.~L.,
  Maiti, T., Ebrahimi, A., Hochrainer, T., Fabritius, H.-O., et~al.,
  \enquote{{DAMASK--The D{\"u}sseldorf Advanced Material Simulation Kit for
  modeling multi-physics crystal plasticity, thermal, and damage phenomena from
  the single crystal up to the component scale},} \emph{Computational Materials
  Science}, Vol. 158, 2019, pp. 420--478.

\bibitem[{Diehl et~al.(2017)Diehl, Groeber, Haase, Molodov, Roters, and
  Raabe}]{diehl2017identifying}
Diehl, M., Groeber, M., Haase, C., Molodov, D.~A., Roters, F., and Raabe, D.,
  \enquote{{Identifying structure--property relationships through DREAM.3D
  representative volume elements and DAMASK crystal plasticity simulations: An
  integrated computational materials engineering approach},} \emph{JOM},
  Vol.~69, No.~5, 2017, pp. 848--855.

\bibitem[{Tran et~al.(2022{\natexlab{a}})Tran, Wildey, and
  Lim}]{tran2022microstructure}
Tran, A., Wildey, T., and Lim, H., \enquote{Microstructure-sensitive
  uncertainty quantification for crystal plasticity finite element constitutive
  models using stochastic collocation method,} \emph{Frontiers in Materials},
  Vol.~9, 2022{\natexlab{a}}, pp. 1--20.

\bibitem[{Xiu and Karniadakis(2002)}]{xiu2002wiener}
Xiu, D., and Karniadakis, G.~E., \enquote{The {W}iener--{A}skey polynomial
  chaos for stochastic differential equations,} \emph{SIAM Journal on
  Scientific Computing}, Vol.~24, No.~2, 2002, pp. 619--644.

\bibitem[{Babu{\v{s}}ka et~al.(2007)Babu{\v{s}}ka, Nobile, and
  Tempone}]{babuvska2007stochastic}
Babu{\v{s}}ka, I., Nobile, F., and Tempone, R., \enquote{A stochastic
  collocation method for elliptic partial differential equations with random
  input data,} \emph{SIAM Journal on Numerical Analysis}, Vol.~45, No.~3, 2007,
  pp. 1005--1034.

\bibitem[{Nobile et~al.(2008)Nobile, Tempone, and Webster}]{nobile2008sparse}
Nobile, F., Tempone, R., and Webster, C.~G., \enquote{A sparse grid stochastic
  collocation method for partial differential equations with random input
  data,} \emph{SIAM Journal on Numerical Analysis}, Vol.~46, No.~5, 2008, pp.
  2309--2345.

\bibitem[{Xiu(2009)}]{xiu2009fast}
Xiu, D., \enquote{Fast numerical methods for stochastic computations: a
  review,} \emph{Communications in computational physics}, Vol.~5, No. 2-4,
  2009, pp. 242--272.

\bibitem[{Novak and Ritter(1996)}]{novak1996high}
Novak, E., and Ritter, K., \enquote{High dimensional integration of smooth
  functions over cubes,} \emph{Numerische Mathematik}, Vol.~75, No.~1, 1996,
  pp. 79--97.

\bibitem[{Novak and Ritter(1997)}]{novak1997curse}
Novak, E., and Ritter, K., \enquote{The curse of dimension and a universal
  method for numerical integration,} \emph{Multivariate approximation and
  splines}, Springer, 1997, pp. 177--187.

\bibitem[{Novak and Ritter(1999)}]{novak1999simple}
Novak, E., and Ritter, K., \enquote{Simple cubature formulas with high
  polynomial exactness,} \emph{Constructive approximation}, Vol.~15, No.~4,
  1999, pp. 499--522.

\bibitem[{Barthelmann et~al.(2000)Barthelmann, Novak, and
  Ritter}]{barthelmann2000high}
Barthelmann, V., Novak, E., and Ritter, K., \enquote{High dimensional
  polynomial interpolation on sparse grids,} \emph{Advances in Computational
  Mathematics}, Vol.~12, No.~4, 2000, pp. 273--288.

\bibitem[{Sudret(2008)}]{sudret2008global}
Sudret, B., \enquote{Global sensitivity analysis using polynomial chaos
  expansions,} \emph{Reliability engineering \& system safety}, Vol.~93, No.~7,
  2008, pp. 964--979.

\bibitem[{Crestaux et~al.(2009)Crestaux, Le~Ma{\i}tre, and
  Martinez}]{crestaux2009polynomial}
Crestaux, T., Le~Ma{\i}tre, O., and Martinez, J.-M., \enquote{Polynomial chaos
  expansion for sensitivity analysis,} \emph{Reliability Engineering \& System
  Safety}, Vol.~94, No.~7, 2009, pp. 1161--1172.

\bibitem[{Saltelli et~al.(2010)Saltelli, Annoni, Azzini, Campolongo, Ratto, and
  Tarantola}]{saltelli2010variance}
Saltelli, A., Annoni, P., Azzini, I., Campolongo, F., Ratto, M., and Tarantola,
  S., \enquote{Variance based sensitivity analysis of model output. Design and
  estimator for the total sensitivity index,} \emph{Computer physics
  communications}, Vol. 181, No.~2, 2010, pp. 259--270.

\bibitem[{Tang et~al.(2010)Tang, Iaccarino, and Eldred}]{tang2010global}
Tang, G., Iaccarino, G., and Eldred, M., \enquote{Global sensitivity analysis
  for stochastic collocation,} \emph{51st AIAA/ASME/ASCE/AHS/ASC Structures,
  Structural Dynamics, and Materials Conference 18th AIAA/ASME/AHS Adaptive
  Structures Conference 12th}, 2010, p. 2922.

\bibitem[{Giles(2008)}]{giles2008multilevel}
Giles, M.~B., \enquote{Multilevel {Monte Carlo} path simulation,}
  \emph{Operations research}, Vol.~56, No.~3, 2008, pp. 607--617.

\bibitem[{Giles(2015)}]{giles2015multilevel}
Giles, M.~B., \enquote{Multilevel {Monte Carlo} methods,} \emph{Acta Numerica},
  Vol.~24, 2015, pp. 259--328.

\bibitem[{Haji-Ali et~al.(2016{\natexlab{a}})Haji-Ali, Nobile, and
  Tempone}]{haji2016multi}
Haji-Ali, A.-L., Nobile, F., and Tempone, R., \enquote{Multi-index {Monte
  Carlo}: when sparsity meets sampling,} \emph{Numerische Mathematik}, Vol.
  132, No.~4, 2016{\natexlab{a}}, pp. 767--806.

\bibitem[{Haji-Ali et~al.(2016{\natexlab{b}})Haji-Ali, Nobile, Tamellini, and
  Tempone}]{haji2016multi2}
Haji-Ali, A.-L., Nobile, F., Tamellini, L., and Tempone, R.,
  \enquote{Multi-index stochastic collocation for random {PDEs},}
  \emph{Computer Methods in Applied Mechanics and Engineering}, Vol. 306,
  2016{\natexlab{b}}, pp. 95--122.

\bibitem[{Haji-Ali et~al.(2016{\natexlab{c}})Haji-Ali, Nobile, Tamellini, and
  Tempone}]{haji2016multi3}
Haji-Ali, A.-L., Nobile, F., Tamellini, L., and Tempone, R.,
  \enquote{Multi-index stochastic collocation convergence rates for random
  {PDEs} with parametric regularity,} \emph{Foundations of Computational
  Mathematics}, Vol.~16, No.~6, 2016{\natexlab{c}}, pp. 1555--1605.

\bibitem[{Benner et~al.(2015)Benner, Gugercin, and Willcox}]{benner2015survey}
Benner, P., Gugercin, S., and Willcox, K., \enquote{A survey of
  projection-based model reduction methods for parametric dynamical systems,}
  \emph{SIAM review}, Vol.~57, No.~4, 2015, pp. 483--531.

\bibitem[{Sirovich(1987)}]{sirovich1987turbulence}
Sirovich, L., \enquote{Turbulence and the dynamics of coherent structures. {I}.
  {C}oherent structures,} \emph{Quarterly of applied mathematics}, Vol.~45,
  No.~3, 1987, pp. 561--571.

\bibitem[{Bunge(2013)}]{bunge2013texture}
Bunge, H.-J., \emph{Texture analysis in materials science: mathematical
  methods}, Elsevier, 2013.

\bibitem[{Kalidindi(2015)}]{kalidindi2015hierarchical}
Kalidindi, S.~R., \emph{Hierarchical materials informatics: novel analytics for
  materials data}, Elsevier, 2015.

\bibitem[{Golub and Van~Loan(2013)}]{golub2013matrix}
Golub, G.~H., and Van~Loan, C.~F., \emph{Matrix computations}, Johns Hopkins
  University Press, 2013.

\bibitem[{Berkooz et~al.(1993)Berkooz, Holmes, and Lumley}]{berkooz1993proper}
Berkooz, G., Holmes, P., and Lumley, J.~L., \enquote{The proper orthogonal
  decomposition in the analysis of turbulent flows,} \emph{Annual review of
  fluid mechanics}, Vol.~25, No.~1, 1993, pp. 539--575.

\bibitem[{Bui-Thanh et~al.(2008)Bui-Thanh, Willcox, and Ghattas}]{bui2008model}
Bui-Thanh, T., Willcox, K., and Ghattas, O., \enquote{Model reduction for
  large-scale systems with high-dimensional parametric input space,} \emph{SIAM
  Journal on Scientific Computing}, Vol.~30, No.~6, 2008, pp. 3270--3288.

\bibitem[{Carlberg et~al.(2017)Carlberg, Barone, and
  Antil}]{carlberg2017galerkin}
Carlberg, K., Barone, M., and Antil, H., \enquote{Galerkin v. least-squares
  {Petrov--Galerkin} projection in nonlinear model reduction,} \emph{Journal of
  Computational Physics}, Vol. 330, 2017, pp. 693--734.

\bibitem[{Tran et~al.(2022{\natexlab{b}})Tran, Eldred, Wildey, McCann, Sun, and
  Visintainer}]{tran2022aphbo}
Tran, A., Eldred, M., Wildey, T., McCann, S., Sun, J., and Visintainer, R.~J.,
  \enquote{{aphBO-2GP-3B}: a budgeted asynchronous parallel multi-acquisition
  functions for constrained {B}ayesian optimization on high-performing
  computing architecture,} \emph{Structural and Multidisciplinary
  Optimization}, Vol.~65, No.~4, 2022{\natexlab{b}}, pp. 1--45.

\bibitem[{Tran et~al.(2019)Tran, Sun, Furlan, Pagalthivarthi, Visintainer, and
  Wang}]{tran2019pbo}
Tran, A., Sun, J., Furlan, J.~M., Pagalthivarthi, K.~V., Visintainer, R.~J.,
  and Wang, Y., \enquote{{pBO-2GP-3B: A batch parallel known/unknown
  constrained Bayesian optimization with feasibility classification and its
  applications in computational fluid dynamics},} \emph{Computer Methods in
  Applied Mechanics and Engineering}, Vol. 347, 2019, pp. 827--852.

\bibitem[{Tran et~al.(2020)Tran, Tranchida, Wildey, and
  Thompson}]{tran2020multi}
Tran, A., Tranchida, J., Wildey, T., and Thompson, A.~P.,
  \enquote{Multi-fidelity machine-learning with uncertainty quantification and
  {B}ayesian optimization for materials design: {A}pplication to ternary random
  alloys,} \emph{The Journal of Chemical Physics}, Vol. 153, 2020, p. 074705.

\bibitem[{Tran and Wildey(2020)}]{tran2020solving}
Tran, A., and Wildey, T., \enquote{Solving stochastic inverse problems for
  property-structure linkages using data-consistent inversion and machine
  learning,} \emph{JOM}, Vol.~73, 2020, pp. 72--89.

\bibitem[{Butler et~al.(2018{\natexlab{a}})Butler, Jakeman, and
  Wildey}]{butler2018convergence}
Butler, T., Jakeman, J., and Wildey, T., \enquote{Convergence of Probability
  Densities Using Approximate Models for Forward and Inverse Problems in
  Uncertainty Quantification,} \emph{SIAM Journal on Scientific Computing},
  Vol.~40, No.~5, 2018{\natexlab{a}}, pp. A3523--A3548.

\bibitem[{Butler et~al.(2018{\natexlab{b}})Butler, Jakeman, and
  Wildey}]{butler2018combining}
Butler, T., Jakeman, J., and Wildey, T., \enquote{Combining push-forward
  measures and {Bayes}' rule to construct consistent solutions to stochastic
  inverse problems,} \emph{SIAM Journal on Scientific Computing}, Vol.~40,
  No.~2, 2018{\natexlab{b}}, pp. A984--A1011.

\bibitem[{Butler et~al.(2020)Butler, Wildey, and Yen}]{butler2020stochastic}
Butler, T., Wildey, T., and Yen, T.~Y., \enquote{Data-consistent inversion for
  stochastic input-to-output maps,} \emph{Inverse Problems}, 2020.

\bibitem[{Wong et~al.(2016)Wong, Madivala, Prahl, Roters, and
  Raabe}]{wong2016crystal}
Wong, S.~L., Madivala, M., Prahl, U., Roters, F., and Raabe, D., \enquote{A
  crystal plasticity model for twinning-and transformation-induced plasticity,}
  \emph{Acta Materialia}, Vol. 118, 2016, pp. 140--151.

\end{thebibliography}

\end{document}